\documentclass{aa}
\newif\iffigure
\figurefalse
\figuretrue

\DeclareRobustCommand{\erase}{\bgroup\markoverwith{\textcolor{red}{\rule[.5ex]{2pt}{2pt}}}\ULon}

\usepackage[switch]{lineno}
\usepackage{graphicx,natbib,url,twoopt}
\usepackage[varg]{txfonts}           %% A&A font choice
\usepackage{hyperref}              %% for pdflatex
\usepackage{pdfcomment}              %% for popup acronym meanings
\usepackage{acronym}                 %% for popup acronym meanings
\usepackage{ulem}                    % 取り消し線を引く
\usepackage{cases}
\usepackage{empheq}                  %Figs. 1,2 
\usepackage{here}
\usepackage{bm}
\usepackage{tabularx}
\usepackage{makecell} % for table contents
\usepackage[version=3]{mhchem}
\hypersetup{% hyperref option
 colorlinks=true,%
 linkcolor=blue,
 citecolor=blue,
}

\usepackage[
  draft,
  authormarkup=none,
  commandnameprefix=ifneeded
]{changes}% finalをdraftにするとハイライト表示
\definechangesauthor[color=black]{R1}
\definechangesauthor[color=black]{R2}
\definechangesauthor[color=black]{R3}
\definechangesauthor[color=black]{R4} % after submission

\usepackage{framed}
\usepackage{mdframed}

\newcommand*\patchAmsMathEnvironmentForLineno[1]{
  \expandafter\let\csname old#1\expandafter\endcsname\csname #1\endcsname
  \expandafter\let\csname oldend#1\expandafter\endcsname\csname end#1\endcsname
  \renewenvironment{#1}
     {\linenomath\csname old#1\endcsname}
     {\csname oldend#1\endcsname\endlinenomath}}
\newcommand*\patchBothAmsMathEnvironmentsForLineno[1]{
  \patchAmsMathEnvironmentForLineno{#1}
  \patchAmsMathEnvironmentForLineno{#1*}}
\AtBeginDocument{
\patchBothAmsMathEnvironmentsForLineno{equation}
\patchBothAmsMathEnvironmentsForLineno{align}
\patchBothAmsMathEnvironmentsForLineno{flalign}
\patchBothAmsMathEnvironmentsForLineno{alignat}
\patchBothAmsMathEnvironmentsForLineno{gather}
\patchBothAmsMathEnvironmentsForLineno{multline}
}

\defcitealias{kuwahara2026multi_iso}{KL26b}
\bibpunct{(}{)}{;}{a}{}{,}    %% natbib cite format used by A&A and ApJ

\makeatletter
\newcommand{\bibnote}[2]{\global\@namedef{#1note}{#2}}
\newcommand{\biblink}[2]{\global\@namedef{#1link}{#2}}

\newcommand{\Equref}[1]{Eq.~\ref{#1}}

\makeatother

\makeatletter
 \newcommandtwoopt{\citeads}[3][][]{%
   \nonstopmode%              %% fix to not stop at error message in latex
   \href{http://adsabs.harvard.edu/abs/#3}%
        {\def\hyper@linkstart##1##2{}%
         \let\hyper@linkend\@empty\citealp[#1][#2]{#3}}%   %% Rutten, 2000
   \biblink{#3}{\href{http://adsabs.harvard.edu/abs/#3}{ADS}}%
   \errorstopmode}            %% fix to resume stopping at error messages 
 \newcommandtwoopt{\citepads}[3][][]{%
   \nonstopmode%              %% fix to not stop at error message in latex
   \href{http://adsabs.harvard.edu/abs/#3}%
        {\def\hyper@linkstart##1##2{}%
         \let\hyper@linkend\@empty\citep[#1][#2]{#3}}%     %% (Rutten 2000)
   \biblink{#3}{\href{http://adsabs.harvard.edu/abs/#3}{ADS}}%
   \errorstopmode}            %% fix to resume stopping at error messages
 \newcommandtwoopt{\citetads}[3][][]{%
   \nonstopmode%              %% fix to not stop at error message in latex
   \href{http://adsabs.harvard.edu/abs/#3}%
        {\def\hyper@linkstart##1##2{}%
         \let\hyper@linkend\@empty\citet[#1][#2]{#3}}%     %% Rutten (2000)
   \biblink{#3}{\href{http://adsabs.harvard.edu/abs/#3}{ADS}}%
   \errorstopmode}            %% fix to resume stopping at error messages 
 \newcommandtwoopt{\citeyearads}[3][][]{%
   \nonstopmode%              %% fix to not stop at error message in latex
   \href{http://adsabs.harvard.edu/abs/#3}%
        {\def\hyper@linkstart##1##2{}%
         \let\hyper@linkend\@empty\citeyear[#1][#2]{#3}}%  %% 2000
   \biblink{#3}{\href{http://adsabs.harvard.edu/abs/#3}{ADS}}%
   \errorstopmode}            %% fix to resume stopping at error messages 

\renewcommand{\@biblabel}[1]{[#1]} % これは番号の出力形式（任意）

\newacro{ADS}{Astrophysics Data System}
\newacro{NLTE}{non-local thermodynamic equilibrium}
\newacro{NASA}{National Aeronautics and Space Administration}

\begin{document}

%\linenumbers
\authorrunning{Kuwahara and Lambrechts}
\titlerunning{Dust transport in envelopes of disk-embedded planets II}

   \title{Dust transport in envelopes of disk-embedded planets}
   \subtitle{II. Fully convective envelopes}
      \author{Ayumu Kuwahara\inst{1} 
          \thanks{\email{ayumu.kuwahara@sund.ku.dk}} 
          \and Michiel Lambrechts\inst{1}}

   \institute{Center for Star and Planet Formation, Globe Institute, University of Copenhagen, Øster Voldgade 5-7, 1350 Copenhagen, Denmark
         }

   \date{Received September XXX; accepted YYY}

  \abstract
    {We perform three-dimensional multifluid simulations of gas and dust to quantify how convection reshapes the spatial distribution of dust grains inside the envelopes of disk-embedded planets. Building on the first paper of this series\added[id=R3]{, which explored convectively stable envelopes, we here consider envelopes in which convection is sustained by the accretion luminosity.}
    The \added[id=R3]{dust-to-gas ratio inside the planetary envelope} is set primarily by the degree of dynamical isolation from the surrounding disk. 
    When convection extends beyond the Bondi radius, the envelope remains connected to the disk flow and maintains a dust-rich state through continuous material exchange. 
    Conversely, when an inner convective layer is separated from an outer recycling layer, the recycling flow filters incoming solids. The dust distribution inside the convective layer is then controlled by the competition between convective stirring and gravitational settling. 
    This yields two regimes: a settling-dominated regime, in which large grains (typically $s\gtrsim0.1$ cm) fall at the terminal velocity and the envelope becomes dust depleted, and a convection-dominated regime, in which smaller grains are trapped in convective circulation and the envelope retains its dust.
    These results imply that the dust and volatile content of planetary envelopes is fundamentally controlled by whether convection enables sustained exchange with the disk or instead operates in an isolated interior.
    \added[id=R3]{The delivery and release of volatile species from accreting solids at the envelope sublimation front, together with the efficiency of convective mixing, determine how this material is distributed  between the core and envelope. 
    Our findings imply that super-Earths and mini-Neptunes inside the water snowline have volatile-depleted cores and volatile-rich envelopes, while  planets at wider orbits can have a wide diversity of envelope compositions from depleted to enriched. 
    This compositional diversity for planetary  envelopes appears to be consistent with the large diversity in observed atmospheric compositions of mini-Neptunes.}
    }
    \keywords{Hydrodynamics --
                Planet-disk interactions --
                Planets and satellites: atmospheres --
                Protoplanetary disks}

   \maketitle

%---------------------------------------------------------
%---------------------------------------------------------
%---------------------------------------------------------
%%%%%%%%%%%%%%%%%%%%%%%%%%%%%%%%%%%%%%%%%%%%%%%%%%%%%%%%%%
%%%%%%%%%%%%%%%%%%%%%%%%%%%%%%%%%%%%%%%%%%%%%%%%%%%%%%%%%%
%%%%%%%%%%%%%%%%%%%%%%%%%%%%%%%%%%%%%%%%%%%%%%%%%%%%%%%%%%

\section{Introduction}\label{sec:Introduction}
Low-mass protoplanets ($\lesssim10\,M_\oplus$; Earth masses) embedded in protoplanetary disks grow by accreting millimeter- to centimeter-sized dust grains (pebble accretion; \citealt{Ormel:2010,Lambrechts:2012}). 
The dust grains can enter the envelope and contribute volatile and refractory material to the core--envelope system.
The accretion of these solids is a key process in setting the primordial composition of the planetary core and its envelope \citep[][]{Sato:2016,Ida:2019,johansen2021pebble,johansen2023anatomy,wang2023atmospheric,steinmeyer2023sublimation}. 
However, pebble accretion does not directly translate into material delivery to the planet because the fate of solids depends on how they are transported within the planetary envelope.

Dust dynamics on envelope scales are largely inherited from gas dynamics because dust grains are aerodynamically coupled to the gas \added[id=R3]{\citep{krapp20223d,krapp2024thermodynamic}}. 
Three-dimensional (3D) hydrodynamical simulations have shown that the gas flow around an embedded planet is strongly perturbed, giving rise to development of a recycling layer in the outer part of the envelope, in which disk gas enters at high latitudes and exits near the midplane \citep{Ormel:2015b,Fung:2015,Kurokawa:2018,Kuwahara:2019}. 
Such flow structures deviate substantially from the commonly adopted assumptions of a prescribed gas velocity field (e.g., unperturbed Keplerian shear) or a 1D static envelope. 
These results highlight the necessity of following the coevolution of gas and dust to understand dust transport and accretion within planetary envelopes \citep{Kuwahara:2020a,Kuwahara:2020b,okamura2021growth}.
Whether solids ultimately contribute to the planetary core and the envelope therefore depends on how dust is transported, retained, or recycled within the envelope, making the resulting spatial distribution of dust a critical quantity for determining the efficiency of solid accretion and the resulting planetary composition.

In the first paper of this series \citep[][hereafter \citetalias{kuwahara2026multi_iso}]{kuwahara2026multi_iso}, we performed multifluid simulations of gas and dust in a local frame that corotated with a planet to investigate the dust transport within an envelope. 
Assuming a nearly isothermal convectively stable envelope, \citetalias{kuwahara2026multi_iso} found that small grains ($\lesssim1$~cm) remain entrained in the recycling flow and are largely prevented from entering the envelope, whereas larger grains penetrate the envelope, but rapidly settle onto the core along the midplane. 
As a consequence, the convectively stable envelope becomes strongly dust depleted.

However, planetary envelopes are inherently heated by the accretion of solids, generating an intrinsic luminosity \added[id=R3]{\citep{rafikov2006atmospheres,Lambrechts:2014,piso2014minimum}}. 
Depending on the resulting temperature gradient, a convective layer can develop within the envelope \citep{Lambrechts:2017,zhu2021global,KL26}. 
\added[id=R3]{\cite{KL26} explored the conditions that drive envelope convection in the regime of super-Earths to mini-Netpunes as function of accretion luminosity and different orbital locations (effective cooling times).
For efficient cooling, when the effective cooling time is comparable to or shorter than the orbital time of the planet, the envelope develops a radiative layer that is largely isolated from the surrounding disk gas flows. 
As cooling becomes less efficient, convection emerges and eventually connects the envelope to the disk, leading to a non-isolated fully convective state \citep{KL26}.}
Convective motions are expected to fundamentally alter dust transport and retention \citep{Popovas:2018a,ali2020limits,johansen2020transport}, potentially invalidating the picture of dust depletion that was established for convectively stable envelopes in \citetalias{kuwahara2026multi_iso}.

In this work, we therefore focus on nearly adiabatic convectively unstable envelopes. 
As in \citetalias{kuwahara2026multi_iso}, we perform 3D multifluid simulations of gas and dust in a local frame that corotates with a planet, but we now include an accretion luminosity. 
\added[id=R4]{The goal of \citetalias{kuwahara2026multi_iso} and this study is not to construct fully self-consistent thermal models of planetary envelopes. 
Instead, we consider idealized end-member envelope structures to isolate how the envelope dynamics affects dust transport.}
The paper is organized as follows. 
Section~\ref{sec:Numerical method} describes the numerical setup of our 3D multifluid simulations. 
Section~\ref{sec:Numerical results} demonstrates that convective envelopes can either retain or lose dust, depending on the flow structure. 
In Sect.~\ref{sec:1D analytic model} we extend the analytic formulae developed in \citetalias{kuwahara2026multi_iso} to the convective regime explored in this work.
Section~\ref{sec:Discussions} places our results in the context of previous work and discusses their implications for planet formation and exoplanet observations. 
We summarize our findings in Sect.~\ref{sec:Conclusions}.%---------------------------------------------------------

\begin{table*}[htbp]
\caption{Parameters of the hydrodynamical simulations.}
%\centering
\resizebox{\textwidth}{!}{
\begin{tabular}{lcccccccccc}\hline\hline
     & $m=R_{\rm B}/H_{\rm g,0}$ & St or $s$ & $\beta$-profile & $\beta_0$ & $\beta_{\rm min}$ & $\beta_{\rm max}$ &  Resolution & $t_{\rm end}$ [$\Omega_0^{-1}$] & $r_{\rm in}$ [$R_{\rm B}$] & $r_{\rm out}$ [$R_{\rm B}$]  \\ \hline
     Fiducial runs (fixed St) & $0.1$ & $10^{-4},\,10^{-3},\,10^{-2}$ & Constant & $10^3$ & - & - & $(N_r,\,N_\theta,\,N_\phi)=(256,\,32,\,128)$ & 100 & $0.1$ & $10$  \\
                          & $0.1$ & $10^{-4},\,10^{-3},\,10^{-2}$ & Variable & - & $10^0$ & $10^3$ & $(N_r,\,N_\theta,\,N_\phi)=(256,\,32,\,128)$ & 100 & $0.1$ & $10$  \\\hline
     Fixed dust size runs & $0.1$ & 0.01 cm, 0.1 cm, \added[id=R3]{1 cm} & Constant & $10^3$ & - & -  & $(N_r,\,N_\theta,\,N_\phi)=(256,\,32,\,128)$ & 100 & $0.1$ & $10$\\
                          & $0.1$ & 0.01 cm, 0.1 cm, \added[id=R3]{1 cm} & Variable & - & $10^0$ & $10^3$ & $(N_r,\,N_\theta,\,N_\phi)=(256,\,32,\,128)$ & 100 & $0.1$ & $10$  \\\hline
     \makecell[l]{\added[id=R3]{Constant low-$\beta$ run} \\ (direct comparison to \citetalias{kuwahara2026multi_iso})} & $0.1$ & $10^{-3}$ & Constant & $10^0$ & - & - & $(N_r,\,N_\theta,\,N_\phi)=(256,\,32,\,128)$ & 100 & $0.1$ & $10$  \\\hline
     Convergence tests   & $0.1$ & $10^{-3}$ & Constant & $10^3$ & - & - & $(N_r,\,N_\theta,\,N_\phi)=(384,\,48,\,192)$ & 50 & $0.1$ & $10$ \\
                         & $0.1$ & $10^{-3}$ & Variable & - & $10^0$ & $10^3$ & $(N_r,\,N_\theta,\,N_\phi)=(384,\,48,\,192)$ & 50 & $0.1$ & $10$ \\
                         & $0.1$ & $10^{-3}$ & Constant & $10^3$ & - & - & $(N_r,\,N_\theta,\,N_\phi)=(256,\,32,\,128)$ & 50 & $0.05$ & $10$ \\
     \hline
\end{tabular}
}
\tablefoot{The columns list the dimensionless thermal mass, the Stokes number or the dust size, the dimensionless cooling time profile, the dimensionless cooling time fixed throughout the domain, the minimum value of the radially varying cooling time, the maximum value of the radially varying cooling time, the resolution, the calculation time, the size of the inner boundary, and the size of the outer boundary.}
\label{tab:hydro simulations}
\end{table*}

%---------------------------------------------------------
\section{Numerical methods}\label{sec:Numerical method}

We simulated gas and dust dynamics around a planet embedded in a non-self-gravitating disk using the code Athena++ with the multifluid dust module \citep{stone2020athena++,HuangBai2022}. 
While \citetalias{kuwahara2026multi_iso} conducted simulations in 2D and 3D, all runs in this study were only performed in 3D spherical polar coordinates centered on a planet, where $r$ is the distance from the planet, $\theta$ is the polar angle, and $\phi$ is the azimuth angle. 
Unless noted otherwise, we adopted the same numerical schemes and units as in \citetalias{kuwahara2026multi_iso}.

\subsection{Governing equations}
We assumed that the gas is a compressible, inviscid, non-self-gravitating fluid and the dust is a pressureless fluid.
We solved the same continuity and momentum equations for gas and dust as in \citetalias{kuwahara2026multi_iso} (their Eqs.~1–5).
We only recall the dust momentum equation and summarize the modifications to the gas energy equation,
\begin{align}
    &\frac{\partial (\rho_{\rm d}\bm{v}_{\rm d})}{\partial t}
    +\nabla\cdot(\rho_{\rm d}\bm{v}_{\rm d}\bm{v}_{\rm d})
    =\rho_{\rm d}\bm{f}_{\rm src}
    +\rho_{\rm d}\frac{\bm{v}_{\rm g}-\bm{v}_{\rm d}}{t_{\rm s}},\label{eq:dust momentum eq}\\
    &\frac{\partial E}{\partial t}
    +\nabla\cdot\left[(E+p)\bm{v}_{\rm g}\right]
    =\rho_{\rm g}\bm{v}_{\rm g}\cdot\bm{f}_{\rm src}
    +Q_{\rm cool}+Q_{\rm heat}.\label{eq:energy eq}
\end{align}
Here, $\rho$ is the density\footnote{This is identical to $\rho^{\rm 3D}$ used in \citetalias{kuwahara2026multi_iso}.}, $\bm{v}$ the velocity, and $p$ the pressure; the subscripts g and d denote gas and dust, respectively. The source term $\bm{f}_{\rm src}$ contains the planetary gravity and the Coriolis and tidal forces and is identical to that in \citetalias{kuwahara2026multi_iso}.
The total energy density is given by $E=e+\rho_{\rm g} v_{\rm g}^2/2$ and the internal energy density by $e=p/(\gamma-1)$, with $\gamma=1.43$ being the adiabatic index.
We included the thermal relaxation term as in \citetalias{kuwahara2026multi_iso}, $Q_{\rm cool}$, and we newly added the accretion heating term, $Q_{\rm heat}$, on the right-hand side of Eq.~\ref{eq:energy eq}, which is described in Sect.~\ref{sec:Cooling and heating terms}.

Equation~\ref{eq:dust momentum eq} includes the gas drag force on its right-hand side, with the stopping time given by \citepalias[Eqs.~11 and 14 of][]{kuwahara2026multi_iso}
\begin{align}
t_{\rm s}\equiv \frac{m_{\rm p}u}{|\bm{F}_{\rm drag}|},\quad \bm{F}_{\rm drag}=-\frac{C_{\rm D}}{2}\pi s^2\rho_{\rm g}u\bm{u}\,.\label{eq:stopping time}
\end{align}
\added[id=R2]{Here, $m_{\rm p}=4\pi s^3\rho_\bullet/3$ is the particle mass, $\rho_\bullet=3\,\mathrm{g\,cm^{-3}}$ is the internal density of the dust, $s$ is the dust size, and $C_{\rm D}$ is the drag coefficient, $\bm{u}\equiv \bm{v}_{\rm d}-\bm{v}_{\rm g}$ and $u\equiv |\bm{u}|$. 
The drag coefficient was evaluated depending on the dust size, the mean free path of the gas, $l_{\rm mfp}$, and the particle Reynolds number, ${\rm Re}_{\rm p}$. 
The mean free path is given by $l_{\rm mfp}=\mu m_{\rm H}/(\rho_{\rm g}\sigma_{\rm mol})$, with $\mu,\,m_{\rm H},$ and $\sigma_{\rm mol}$  the mean molecular weight, $\mu=2.34$, the mass of the proton, and the molecular collision cross section, $\sigma_{\rm mol}=2\times10^{-15}\,\text{cm}^2$ \citep[e.g.,][]{Nakagawa:1986}.
The particle Reynolds number is defined as $\mathrm{Re}_{\rm p}=2su/\nu$, with $\nu=l_{\rm mfp}c_{\rm s}(r)/2$ the viscosity, and $c_{\rm s}(r)$ is the local sound speed.}\added[id=R2]{
For $s<9l_{\rm mfp}/4$, we adopted the Epstein drag law,
\begin{align}
    C_{\rm D}=\frac{8c_{\rm s}}{3u}.\label{eq:Epstein drag coefficient}
\end{align}
For larger particles, $s\ge 9l_{\rm mfp}/4$, we used \citep{Weidenschilling:1977}
\begin{align}
    C_{\rm D}=
    \begin{cases}
        24\,\mathrm{Re}_{\rm p}^{-1} & (\mathrm{Re}_{\rm p}<1),\\
        24\,\mathrm{Re}_{\rm p}^{-0.6} & (1<\mathrm{Re}_{\rm p}<800),\\
        0.44 & (\mathrm{Re}_{\rm p}>800).\label{eq:nonlinear drag coefficient}
    \end{cases}
\end{align}}As in \citetalias{kuwahara2026multi_iso}, we neglected dust diffusion and back-reaction of dust on the gas.

The gravitational potential was implemented with a force-free smoothing at the inner boundary \citepalias[Eqs.~7 and 8 of][]{kuwahara2026multi_iso},
\begin{empheq}[left = {\Phi_{\rm p}=\empheqlbrace \,}]{alignat = 2}
    &-\frac{GM_{\rm p}}{r}\, \frac{(r-r_{\rm in})^2}{(r-r_{\rm in})^2+r_{\rm sm}^2}\, f_{\rm inj}
    &&\quad\text{(for the gas)},\\
    &-\frac{GM_{\rm p}}{r}\, f_{\rm inj}
    &&\quad\text{(for the dust)},
\end{empheq}
where $G$ is the gravitational constant, $M_{\rm p}$ is the planet mass, $r_{\rm in}$ is the inner radial boundary, and $r_{\rm sm}=0.1\,r_{\rm in}$ is the smoothing length.  
The factor $f_{\rm inj}=1-\exp[-(t/t_{\rm inj})^2/2]$ gradually inserts the planetary gravity into the disk to prevent shock formation \citep{Ormel:2015a}, with $t_{\rm inj}=5\,\Omega_0^{-1}$ and $\Omega_0$ the orbital frequency at the planet location.

%----------------------------
\subsection{Cooling and heating terms}\label{sec:Cooling and heating terms}
The energy equation includes a cooling term, treated as thermal relaxation, on the right-hand side of \Equref{eq:energy eq}, as in \citetalias{kuwahara2026multi_iso}, and also a heating term that was newly added in this study. 
The cooling was implemented as
\begin{align}
    Q_{\rm cool}=-\frac{e-e_0}{t_{\rm cool}},
\end{align}
where $e_0$ is the initial value of the internal energy density.
The isothermal and adiabatic limits are reached when the cooling timescale approaches $t_{\rm cool}\rightarrow0$ and $t_{\rm cool}\rightarrow\infty$, respectively.

\citetalias{kuwahara2026multi_iso} set $t_{\rm cool}\Omega_0=1$, which yielded a nearly isothermal and convectively stable envelope, satisfying the Schwarzschild criterion for the temperature gradient,
\begin{align}
        \frac{\partial \ln T}{\partial \ln p}<\nabla_{\rm ad}\equiv\frac{\gamma-1}{\gamma}.\label{eq:Schwarzschild criterion}
\end{align}
In contrast, we primarily considered envelopes in the nearly adiabatic regime, $t_{\rm cool}\Omega_0\gg1$, in which the temperature gradient exceeds the adiabatic gradient and the envelope becomes fully convective.
The cooling-time prescription in this study is described in Sect.~\ref{sec:Simulation parameters}.

Accretion heating was included via the luminosity associated with the accretion of solids at a mass rate $\dot{M}$ \citep{rafikov2006atmospheres,Lambrechts:2014}, 
\added[id=R4]{\begin{align}
    L_{\rm acc}\equiv\frac{GM_{\rm p}\dot{M}}{R_{\rm p}}\sim10^{27}\,\mathrm{erg/s}\,\Bigg(\frac{M_{\rm p}}{M_\oplus}\Bigg)^{2/3}\Bigg(\frac{\dot{M}}{10\,M_\oplus\mathrm{/Myr}}\Bigg), \label{eq:Lacc}
\end{align}
where $R_{\rm p}$ is the size of the core.} This was implemented as a source term of the energy equation \citep{zhu2021global,KL26},
\begin{align}
    Q_{\rm heat}=\frac{L_{\rm acc}}{4\pi}\frac{r_{\rm in}}{r^4}.\label{eq:Q heat}
\end{align}
We neglected other heat sources such as envelope contraction, latent heat, and radioactive heating.

% -------------------------------
\subsection{Code units, initial conditions, and fixed parameters}\label{sec:Code units and initial conditions}
We adopted the same code units as in \citetalias{kuwahara2026multi_iso},
\begin{align}
    H_{\rm g,0}=c_{\rm s,0}=\Omega_0=\rho_{\rm g,0}=1,
\end{align}
where $H_{\rm g,0}$ is the gas scale height, $c_{\rm s,0}$ the isothermal sound speed, and $\rho_{\rm g,0}$ is the midplane gas density at the orbital radius of the planet $a_{\rm p}$. 
In contrast to the nearly isothermal assumption in \citetalias{kuwahara2026multi_iso}, the sound speed now depends on the radius, 
\begin{align}
    &c_{\rm s}(r)=\sqrt{\frac{p(r)}{\rho(r)}}\,c_{\rm s,0},
\end{align}
where $p$ and $\rho$ are in code units.

When disk self-gravity is neglected, the planetary mass is normalized by
\begin{align}
    m\equiv\frac{R_{\rm B}}{H_{\rm g,0}}&=\frac{M_{\rm p}}{M_{\rm th}}\simeq0.11\,\Bigg(\frac{M_{\rm p}}{M_\oplus}\Bigg)\Bigg(\frac{M_\ast}{M_\odot}\Bigg)^{-1}\Bigg(\frac{0.03}{h}\Bigg)^3.
\end{align}
Here, $R_{\rm B}=GM_{\rm p}/c_{\rm s,0}^2$ is the Bondi radius, $M_{\rm th}=M_\ast h^3$ is the thermal mass, $M_\ast$ is the stellar mass, $h$ is the disk aspect ratio, and $M_\odot$ is the solar mass.
As in \citetalias{kuwahara2026multi_iso}, we set $m=0.1$, so that the Hill radius in code units is given by $R_{\rm H}/H_{\rm g,0}=(m/3)^{1/3}\simeq0.32$.

The initial gas and dust densities follow vertically stratified Gaussian profiles, identical to those \added[id=R3]{in \citetalias[their {Eq.~17}]{kuwahara2026multi_iso}.}  
We prescribed an initial dust scale height of $H_{\rm d,0}=0.1\,H_{\rm g,0}=R_{\rm B}$ so that the Bondi region was initially dust filled.  
We defined the dust-to-gas ratio by
\begin{align}
    \epsilon\equiv\frac{\rho_{\rm d}}{\rho_{\rm g}},
\end{align}
with $\epsilon_0=0.01$ its initial value. 
The background velocity field was a Keplerian shear flow neglecting the headwind induced by the global pressure gradient \added[id=R3]{\citepalias[Eq.~20 of][]{kuwahara2026multi_iso}.}

For the accretion heating, we parameterized the luminosity via an accretion timescale $t_{\rm acc}=10^5$~yr, which corresponds to a pebble accretion rate of $\dot{M}=10\,M_\oplus/\mathrm{Myr}$ \citep{Lambrechts:2014}. 
This sets the dimensionless luminosity \citep{KL26},
\footnotesize
\begin{align}
    \tilde{L}_{\rm acc}&\equiv\frac{L_{\rm acc}}{c_{\rm s}^2M_{\rm th}\Omega_0},\nonumber\\
    &\simeq1.1\times10^{-5}\,\Bigg(\frac{m}{0.1}\Bigg)^{5/3}\Bigg(\frac{\rho_{\rm p}}{5\,\text{g/cm}^{3}}\Bigg)^{1/3}\Bigg(\frac{M_\ast}{M_\odot}\Bigg)^{-5/6}\Bigg(\frac{a_{\rm p}}{1\,\text{au}}\Bigg)^{5/2}\Bigg(\frac{t_{\rm acc}}{10^5\,\text{yr}}\Bigg)^{-1},
\end{align}
\normalsize
where $\rho_{\rm p}$ is the density of the planet.

% -------------------------------------
\subsection{Simulation parameters}\label{sec:Simulation parameters}
The dimensionless stopping time (Stokes number) is defined by
\begin{align}
    {\rm St}=t_{\rm s}\Omega_0.
\end{align}
In the fiducial runs, we adopted fixed Stokes numbers of ${\rm St}=10^{-4},\,10^{-3},$ and $10^{-2}$ throughout the computational domain.
We also performed fixed-size runs with \added[id=R3]{$s=0.01$, 0.1, and $1$~cm}. 
As in \citetalias{kuwahara2026multi_iso}, dust physics such as growth, fragmentation, erosion, and ablation were not included. 
Table~\ref{tab:hydro simulations} summarizes the parameters of our simulations.

Unlike \citetalias{kuwahara2026multi_iso}, where we enforced a nearly isothermal convectively stable envelope, we primarily considered envelopes that were nearly adiabatic and fully convective. 
\added[id=R4]{As in \citetalias{kuwahara2026multi_iso}, we parameterized the thermal relaxation timescale as the dimensionless quantity $\beta\equiv t_{\rm cool}\Omega$ \citep{Gammie:2001}. 
We performed simulations with a constant-$\beta$ cooling prescription, which is directly comparable to the nearly isothermal envelopes in \citetalias{kuwahara2026multi_iso}, but now including accretion heating.
Following \cite{KL26}, we adopted $\beta=\beta_0=10^0$ and $10^3$ because these values produce convectively stable and fully convective envelopes within the Bondi radius, respectively.
We also explored a radially varying prescription for the cooling time,
\begin{align}
    \beta(r)=\beta_{\rm min}+
      \frac{(\beta_{\rm max}-\beta_{\rm min})}{2}\Bigg[1-\tanh\bigg(\frac{r-r_0}{w}\bigg)\Bigg],\label{eq:variable beta function}
\end{align}
with $r_0=R_{\rm B}$ and $w=0.01\,R_{\rm B}$ (Xu et al., in prep.).
The adopted transition width, $w$, is resolved by a single grid cell, and Eq.~\ref{eq:variable beta function} therefore behaves effectively as a step function.
We set $\beta_{\rm max}=10^3$ and $\beta_{\rm min}=10^0$.
The adopted $\beta$ values were chosen to generate idealized envelope structures and not to represent self-consistently computed cooling times.
The variable-$\beta$ prescription ensures a nearly adiabatic and fully convective layer inside the Bondi radius, but the flow pattern at $r>R_{\rm B}$ differs from a constant $\beta$ prescription (Sect.~\ref{sec:Numerical results}). 
We explored constant and variable cooling regimes because it is a priori not known how dust-opacity-dependent cooling times differ in- and outside the planetary envelope.}

\added[id=R4]{Appendices~\ref{sec:1D envelope structure} and \ref{sec:Envelope cooling time model} provide simple envelope-structure and cooling-time estimates that help interpret the adopted $\beta$ prescriptions.
These estimates are only intended as physical guidance because a fully self-consistent model would require radiative transfer simulations that directly include the grains that act as a source of opacity (Sect.~\ref{sec:Model limitations}).
}
\added[id=R4]{For} a given dust opacity and luminosity (e.g., $\kappa=10^{-1}\,\mathrm{cm^2\,g^{-1}}$ and $L=10^{27}\,\mathrm{erg\,s^{-1}}$), the constant $\beta$ model \added[id=R3]{($\beta_0=10^3$)} is valid for the inner hotter disk regions ($\lesssim$ a few au).
The gas is expected to remain convectively unstable even beyond the Bondi radius, implying that no radiative-convective boundary (RCB) develops within $R_{\rm B}$  \citep{rafikov2006atmospheres}. 
In contrast, the variable-$\beta$ prescription is designed to capture a structure in which the envelope is convectively unstable inside $R_{\rm B}$, but transitions to convectively stable at $r\geq R_{\rm B}$ (the RCB \added[id=R3]{is located} at the Bondi radius). 
Such a configuration is expected to occur farther out in the disk, typically at orbital distances of $\gtrsim$ a few au.

% -------------------------------------
\subsection{Resolution and boundary conditions}
We used a logarithmically spaced grid in radius between $r_{\rm in}=0.1\,R_{\rm B}$ and $r_{\rm out}=10\,R_{\rm B}$, with uniform spacing in $\theta$ and $\phi$.
\added[id=R3]{The size of the inner boundary  corresponds to approximately ten times the physical radius of the planet, which is given by \citepalias[Eq.~21 of][]{kuwahara2026multi_iso}
\begin{align}
    \frac{R_{\rm p}}{H_{\rm g,0}}\simeq1.4\times10^{-3}\,\bigg(\frac{m}{0.1}\bigg)^{1/3}\bigg(\frac{a_{\rm p}}{\text{1 au}}\bigg)^{-1}.\label{eq:core radius}
\end{align}
In} the fiducial runs, the Bondi radius was resolved by 128 radial cells, where we doubled the resolution in the radial direction compared to \citetalias{kuwahara2026multi_iso} (Table~\ref{tab:hydro simulations}). 
We only simulated the upper half of the disk, $\theta\in[0,\pi/2]$. 
The full $2\pi$ range in azimuth was covered. 
At the inner radial boundary, we applied reflecting and outflow boundary conditions for the gas and dust, respectively, while fixing the density and velocity of the gas and dust to their initial values at the outer boundary. 
In the midplane, we used a reflecting boundary, and at the pole, we applied the polar boundary condition of the Athena++, copying conserved variables across the axis \citep{stone2020athena++}. 
Appendix \ref{sec:Convergence tests} provides the convergence tests.

%-----------------------------------
%-----------------------------------

%---------------------------------------------------------
%---------------------------------------------------------

\begin{figure}[tp]
    \centering
    \includegraphics[width=1\linewidth]{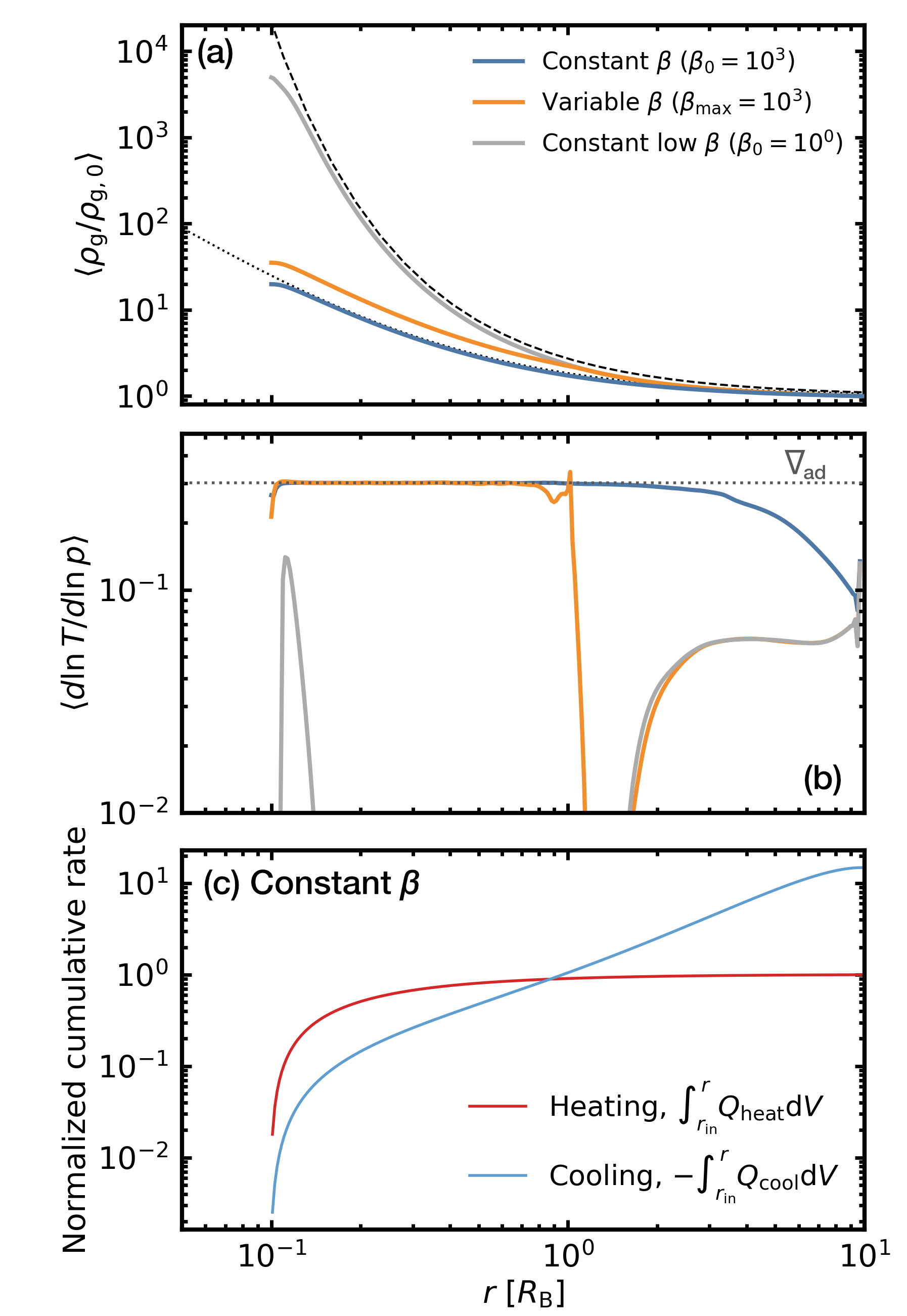}
    \caption{\added[id=R4]{Shell-averaged density and temperature gradient for different $\beta$ prescriptions, and cumulative volume-integrated heating and cooling rates. \textit{Top}: Assumed hydrostatic equilibrium in adiabatic and isothermal limits is shown by the dotted and dashed curves(Eqs.~\ref{eq:rhog ad} and \ref{eq:rho g iso}). \textit{Middle}: Adiabatic gradient (dotted line; Eq.~\ref{eq:Schwarzschild criterion}). \textit{Bottom:} Vertical axis normalized by the volume-integrated heating rate, $\int_{r_{\rm in}}^{r_{\rm out}}Q_{\rm heat}\mathrm{d}V$.}}
    \label{fig:rho_dlnT_dlnp_cumQ}
\end{figure}

\begin{figure}[tp]
    \centering
    \includegraphics[width=1\linewidth]{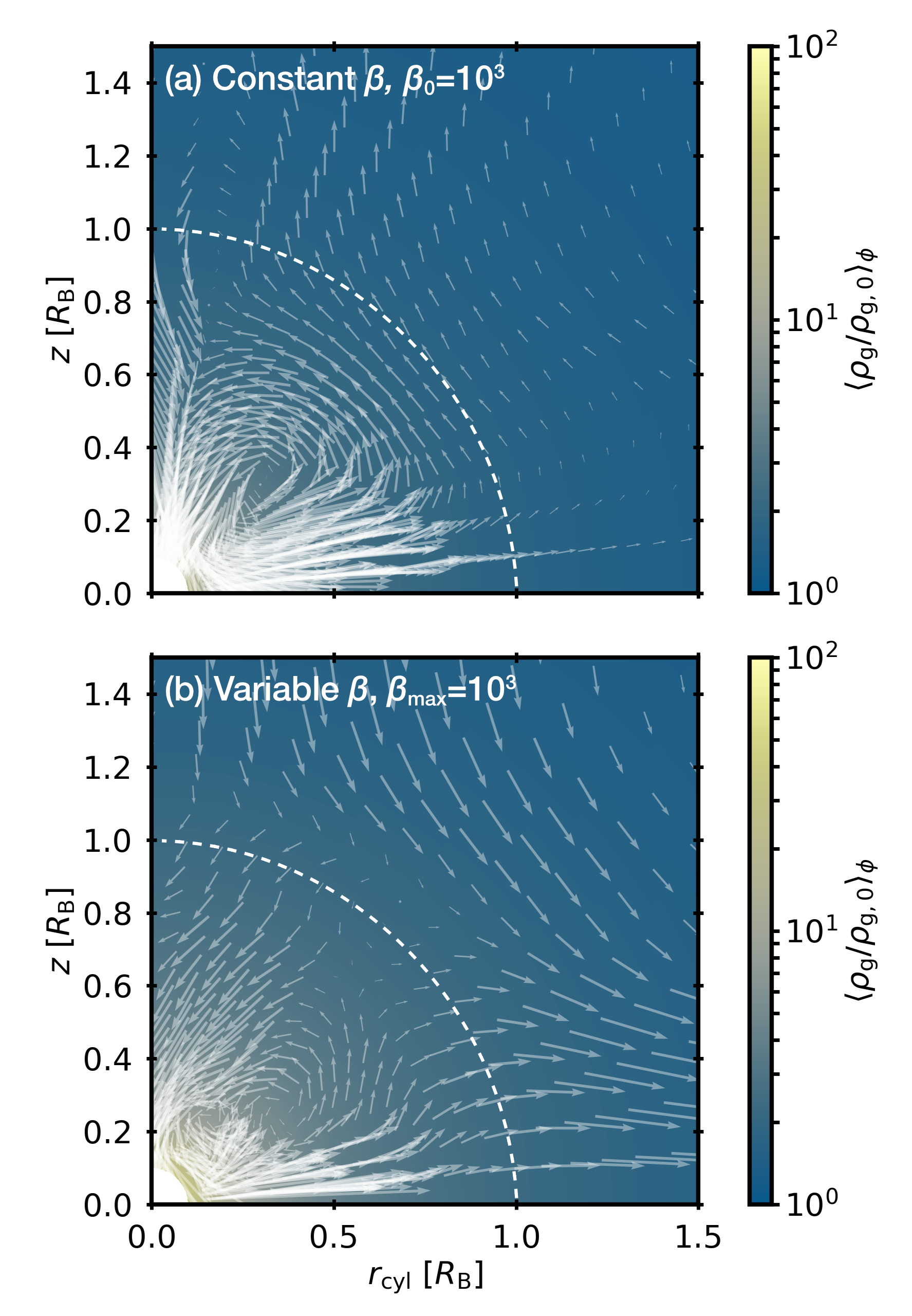}
    \caption{\added[id=R4]{Density and velocity vector field of the gas averaged over the azimuth, $\phi\in[-\pi/2,\,\pi/2]$,} for \added[id=R3]{the two} different $\beta$ prescriptions \added[id=R3]{at $t=93$ (\textit{top}) and $t=86$ (\textit{bottom}).} The dashed white circle marks the Bondi radius.}
    \label{fig:slice_xz_rhog_vertical_b1e+03}
\end{figure}

\begin{figure}[tp]
    \centering
    \includegraphics[width=1\linewidth]{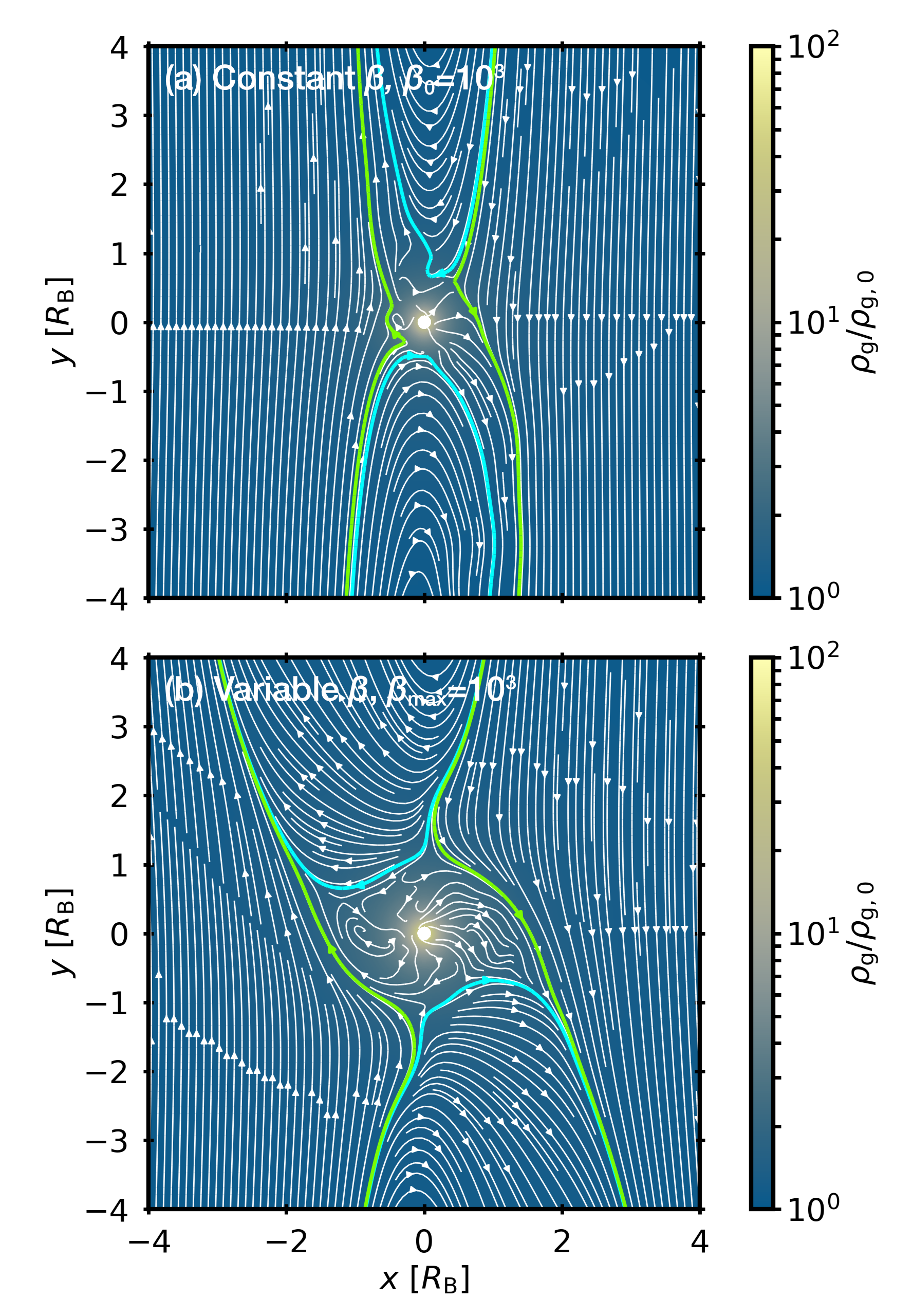}
    \caption{Gas flow on the midplane slice with gas streamlines for \added[id=R3]{the two} different $\beta$ prescriptions \added[id=R3]{at $t=93$ (\textit{top}) and $t=86$ (\textit{bottom}).} The cyan and green curves mark\added[id=R3] the outer horseshoe and inner shear streamlines, respectively.}
    \label{fig:slice_xy_rhog_vertical_b1e+03}
\end{figure}

\begin{figure}[tp]
    \centering
    \includegraphics[width=1\linewidth]{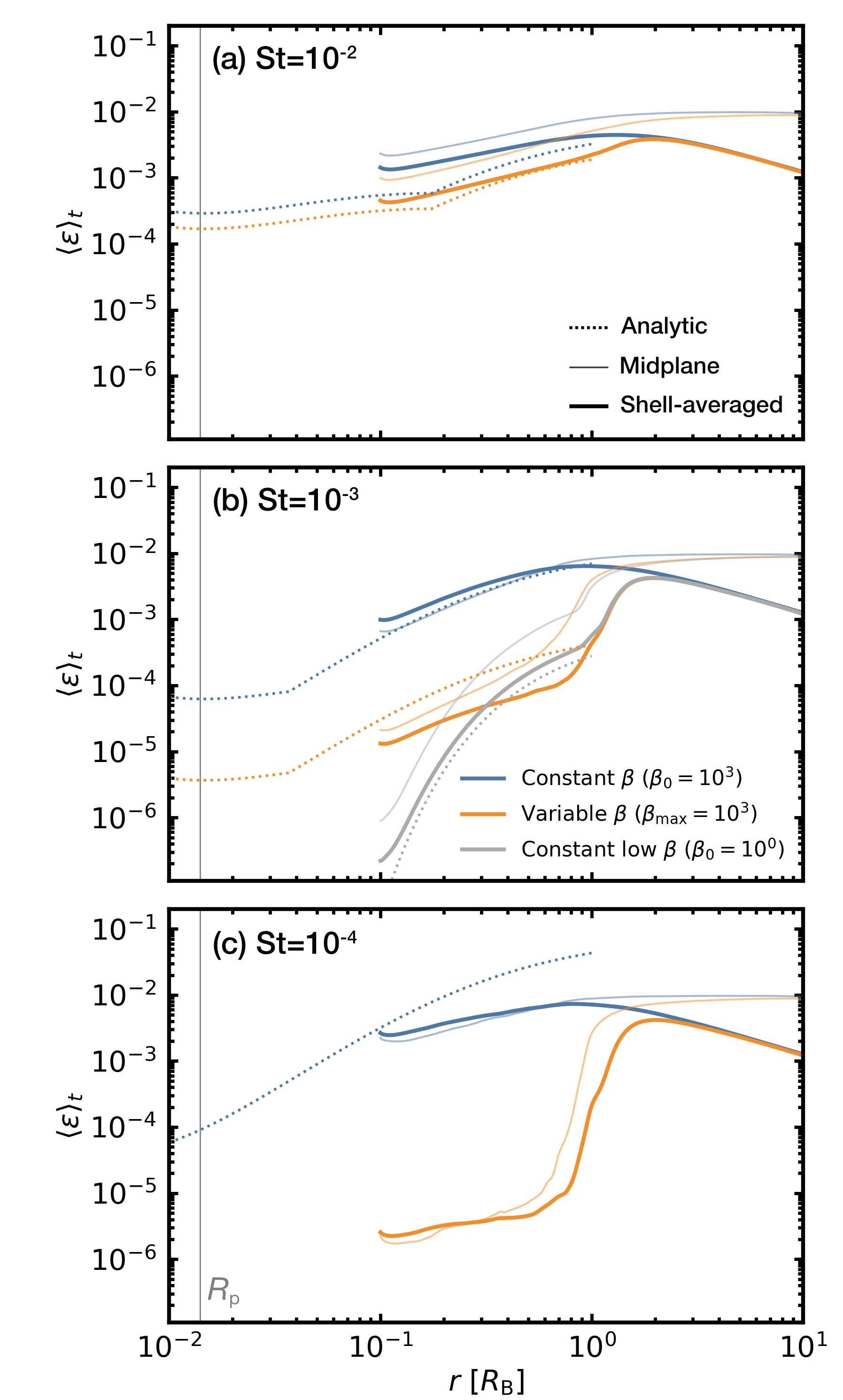}
    \caption{\added[id=R3]{Steady-state} dust-to-gas ratio for different $\beta$ prescriptions and Stokes numbers. The thick and thin solid curves show the \added[id=R3]{shell-averaged} value and the \added[id=R3]{azimuthally averaged} value in the midplane, respectively. \added[id=R4]{These curves are also time-averaged over the last ten orbits for clarity.} The dotted curves are given by Eq.~\ref{eq:epsilon model}, introduced in Sect.~\ref{sec:1D analytic model}. In panel c, the analytic model formally predicts complete dust filtering in the variable-$\beta$ case and is therefore not shown. \added[id=R3]{The vertical line marks the core radius (Eq.~\ref{eq:core radius}).}}
    \label{fig:1dslice_d_to_g_b1e+03}
\end{figure}

\begin{figure}[tp]
    \centering
    \includegraphics[width=1\linewidth]{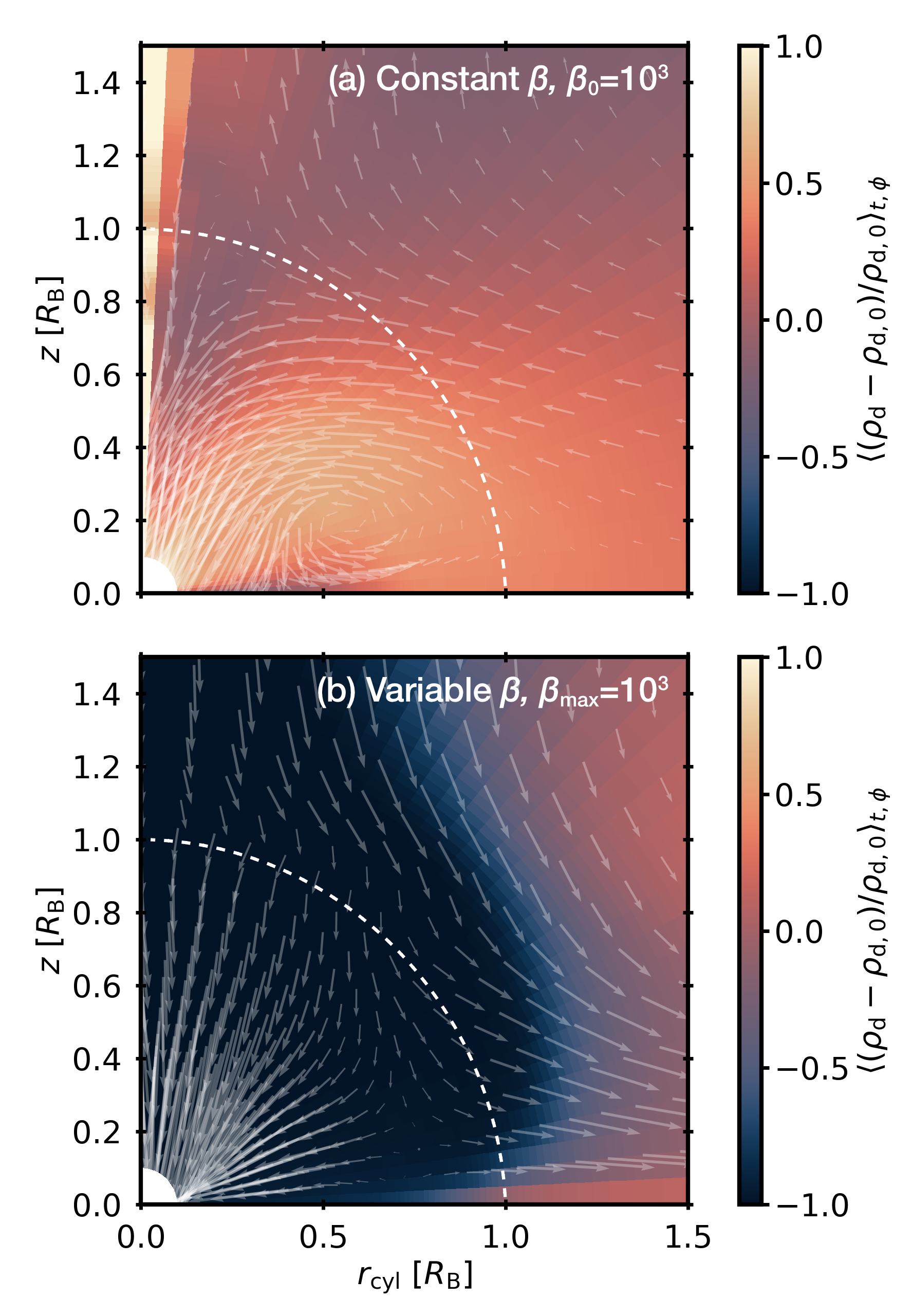}
    \caption{\added[id=R4]{Density and velocity vector field of the dust averaged over the azimuth, $\phi\in[-\pi/2,\,\pi/2]$,} for a fixed Stokes number ${\rm St}=10^{-3}$ \added[id=R3]{that are also \added[id=R4]{time-}averaged over the last ten orbits.}  The background color shows the deviation from the initial dust distribution. Brighter and darker regions indicate dust accumulation and depletion, respectively. \added[id=R3]{A non-isolated convection-dominated envelope (constant $\beta_0=10^3$) retains \added[id=R4]{dust by exchanging material with the disk.}}}
    \label{fig:slice_xz_rhod_b1e+03}
\end{figure}

\begin{figure}[tp]
    \centering
    \includegraphics[width=1\linewidth]{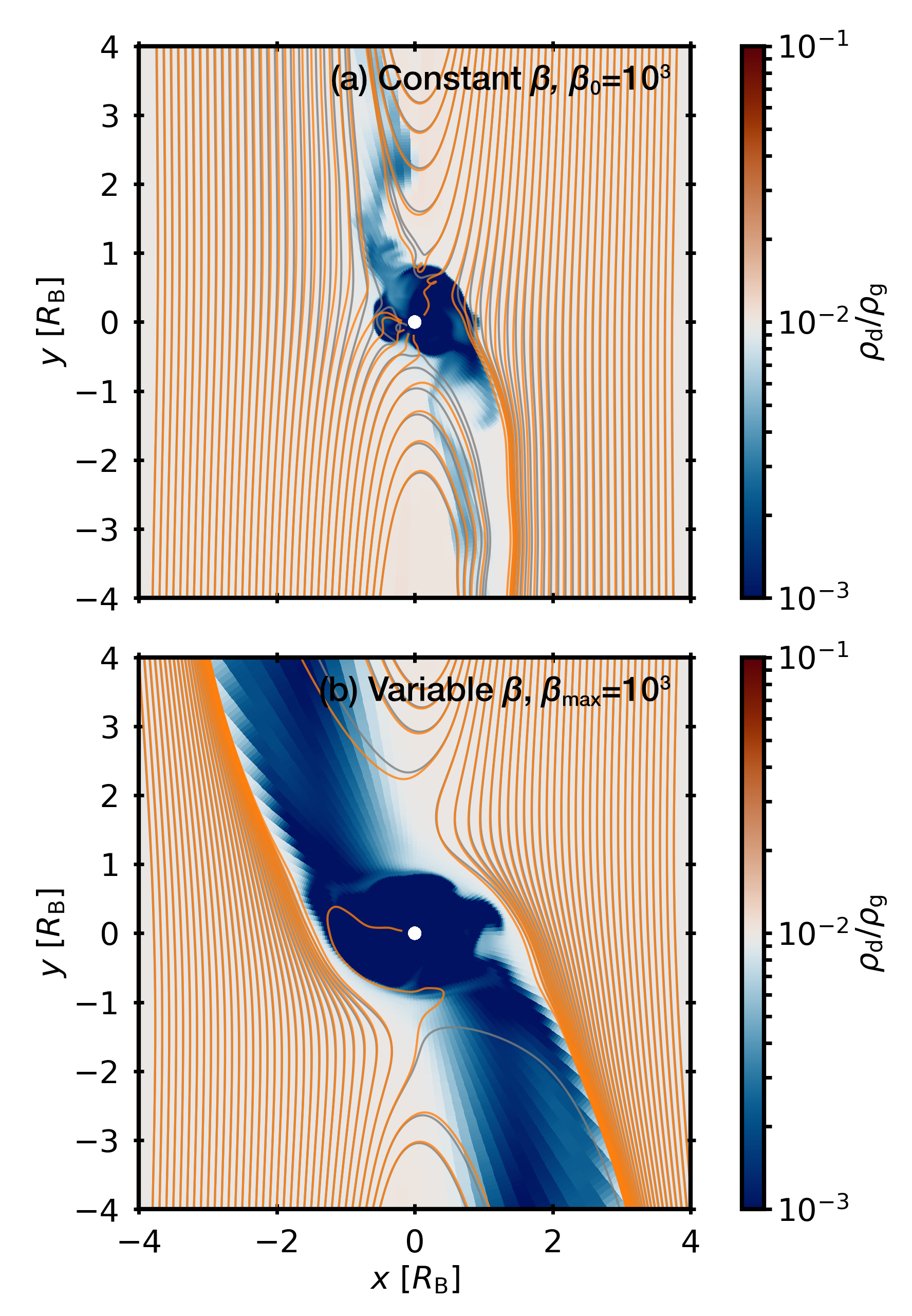}
    \caption{Streamlines in the midplane of gas (gray) and dust (orange) \added[id=R3]{at $t=93$ (\textit{top}) and $t=86$ (\textit{bottom})}, originating from the first and third quadrants. The background color shows the dust-to-gas ratio. We adopted \added[id=R3]{a} fixed Stokes number, ${\rm St}=10^{-3}$.}
    \label{fig:slice_xy_dtog_b1e+03}
\end{figure}

\begin{figure}[tp]
    \centering
    \includegraphics[width=1\linewidth]{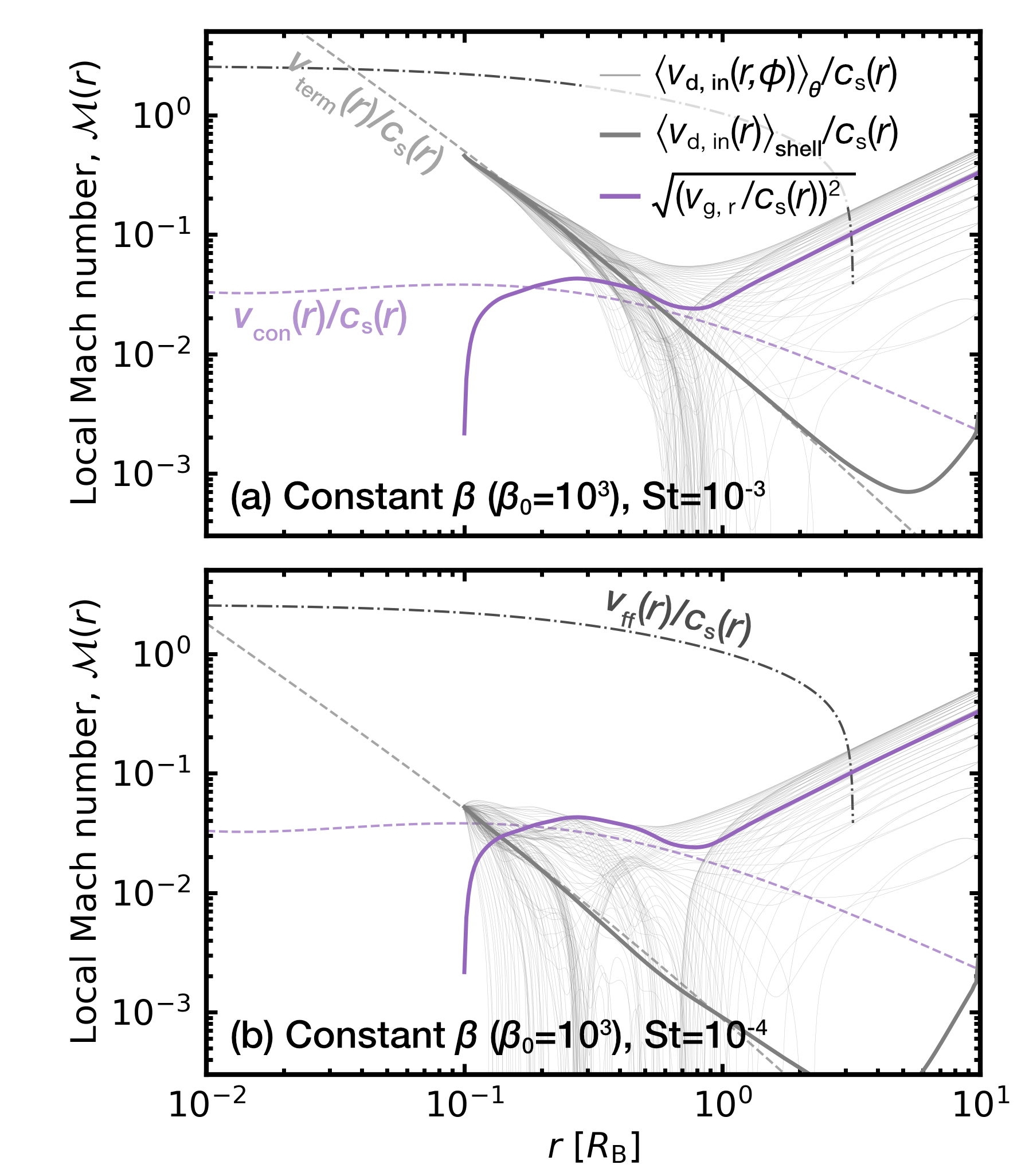}
    \caption{Local Mach numbers of the radially inward dust velocity (solid gray) and the root-mean-square of the radial gas velocity (solid purple). The thin gray curves show the dust velocity at different azimuths \added[id=R4]{averaged over the polar angle, and} the thick gray curve shows the shell-averaged dust velocity. The dashed gray, dot-dashed black, and dashed purple curves are given by Eqs.~\ref{eq:vterm}--\ref{eq:vcon}. Only the constant-$\beta$ case is shown because the variable-$\beta$ run did not differ significantly.}
    \label{fig:1dslice_local_mach_number_vertical}
\end{figure}

%---------------------------------------------------------
%---------------------------------------------------------
\section{Numerical results}\label{sec:Numerical results}
Convective envelopes can either retain or lose dust, depending on whether they remain dynamically connected to the surrounding disk or become effectively isolated \added[id=R3]{\citep{KL26}}.
We first summarize the gas dynamics of convective envelopes and quantify their degree of isolation from the disk flow (Sect.~\ref{sec:Gas dynamics and envelope isolation}).
We then demonstrate how this isolation controls the dust-to-gas ratio \added[id=R3]{within the envelope} (Sect.~\ref{sec:Dust-retaining or dust-depleted envelopes}) before we examine the underlying dust dynamics that cause retention or depletion (Sect.~\ref{sec:Dust dynamics in convective envelopes}).
Finally, we extend the 1D analytic model of \citetalias{kuwahara2026multi_iso} to the convective regime explored in this work (Sect.~\ref{sec:1D analytic model}).

%-------------------------------------------------------------------------------------------
\subsection{Gas dynamics and envelope isolation}\label{sec:Gas dynamics and envelope isolation}
Inefficient cooling (dimensionless cooling time $\beta=10^3$) yields a nearly adiabatic convectively unstable envelope.
Hydrodynamical simulations have extensively studied such convective flows \citep[e.g.,][]{Lambrechts:2017,Popovas:2018b,zhu2021global,KL26}.
We therefore only focused on the aspects that are essential for interpreting our results and for highlighting the differences from our previous study of convectively stable envelopes \citepalias{kuwahara2026multi_iso}.

Figure~\ref{fig:rho_dlnT_dlnp_cumQ} shows radial profiles of the gas density, \added[id=R4]{the temperature gradient, and the cumulative heating and cooling rates}.
The gas density in the constant and variable $\beta$ runs follows the adiabatic limit \citep[e.g.,][Fig.~\ref{fig:rho_dlnT_dlnp_cumQ}a]{rafikov2006atmospheres},
\begin{align}
    \rho_{\rm g}(r)\approx\rho_0\left(1+\nabla_{\rm ad}\frac{R_{\rm B}}
        {\sqrt{r^2+r_{\rm sm}^2}}\right)^{1/(\gamma-1)}\equiv\rho_{\rm g, ad}(r).\label{eq:rhog ad}
\end{align}

\added[id=R4]{In the constant-$\beta$ run ($\beta_0=10^3$)}, ${\rm d}\ln T/{\rm d}\ln p$ follows the adiabatic value even outside the Bondi radius, indicating that the convectively unstable region extends beyond $R_{\rm B}$ (blue curve in Fig.~\ref{fig:rho_dlnT_dlnp_cumQ}b).
In the variable-$\beta$ run ($\beta_{\rm max}=10^3$ and $\beta_{\rm min}=10^0$), the flow is convectively unstable inside $R_{\rm B}$ but becomes convectively stable outside (orange curve in Fig.~\ref{fig:rho_dlnT_dlnp_cumQ}b).

For reference, in the constant low-$\beta$ run ($\beta_0=10^0$), the gas remains convectively stable throughout the domain (gray curve in Fig.~\ref{fig:rho_dlnT_dlnp_cumQ}), \added[id=R3]{and its density follows the isothermal limit,
\begin{align}
    \rho_{\rm g}(r)\approx\rho_{\rm g,0}\exp\Bigg(\frac{R_{\rm B}}{\sqrt{r^2+r_{\rm sm}^2}}\Bigg)\equiv\rho_{\rm g, iso}(r).\label{eq:rho g iso}
\end{align}}\added[id=R4]{Heating is concentrated in the deep envelope, whereas cooling occurs over a broader region, so the two are not locally balanced (Fig.~\ref{fig:rho_dlnT_dlnp_cumQ}c).
The resulting excess energy is transported outward primarily by convection.}

Convective motions are clearly visible in the gas velocity field (Fig.~\ref{fig:slice_xz_rhog_vertical_b1e+03}).
The key distinction for what follows is whether the envelope is dynamically isolated from the disk.
In the constant-$\beta$ run, the envelope remains non-isolated and continuously exchanges gas with the surrounding disk, with no distinct boundary separating gas inside $R_{\rm B}$ from the disk (Fig.~\ref{fig:slice_xz_rhog_vertical_b1e+03}a).
In the variable-$\beta$ run, by contrast, the inner convective layer is effectively separated from the outer recycling layer (Fig.~\ref{fig:slice_xz_rhog_vertical_b1e+03}b).
The recycling flow is characterized by polar inflow and midplane outflow \citep[e.g.,][]{Ormel:2015b}.
Gas entering from high latitudes is prevented from penetrating into $R_{\rm B}$ by a positive entropy gradient, and it then exits through the midplane \added[id=R4]{\citep{Kurokawa:2018}}.

This difference is also evident in midplane slices (Fig.~\ref{fig:slice_xy_rhog_vertical_b1e+03}): the constant-$\beta$ case shows no clear demarcation between the envelope and the background disk flow, whereas the gas within $R_{\rm B}$ is largely decoupled from the surrounding shear and horseshoe streamlines in the variable-$\beta$ case (green and cyan curves).
Because dust dynamics are inherited from gas dynamics, this dynamical isolation plays a central role in setting how much dust can enter and remain in the envelope, as we show next (Sect.~\ref{sec:Dust-retaining or dust-depleted envelopes}).

%-------------------------------------------------------------------------------------------
\subsection{Dust-retaining or dust-depleted envelopes}\label{sec:Dust-retaining or dust-depleted envelopes}
When convection extends beyond the Bondi radius, the envelope is non-isolated and continuously exchanges material with the disk, leading to a dust-retaining envelope.
When the inner convective layer is instead isolated from the outer recycling flow, the envelope is shielded from incoming dust and can become strongly dust depleted, as in \citetalias{kuwahara2026multi_iso}.

Figure~\ref{fig:1dslice_d_to_g_b1e+03}\added[id=R3]{b} shows the radial profile of the dust-to-gas ratio for ${\rm St}=10^{-3}$.
In the constant-$\beta$ (non-isolated) case, the shell-averaged dust-to-gas ratio remains at $\langle\epsilon\rangle \gtrsim 10^{-3}$ throughout the computational domain (blue curve), consistent with sustained dust exchange with the disk.
As the gas density increases toward the planet, $\langle\epsilon\rangle$ gradually decreases inward.
The dust-to-gas ratio measured in the midplane is systematically lower than the shell-averaged value, reflecting vertical redistribution by convection that lofts dust to higher altitudes (Fig.~\ref{fig:slice_xz_rhod_b1e+03}a).
\added[id=R4]{Dust grains can enter the envelope through a polar accretion pathway following a high-altitude horseshoe turn, producing a fluctuating localized enhancement of the dust density along the polar axis (Fig.~\ref{fig:slice_xz_rhod_b1e+03}a).
Although azimuthal averaging can enhance the apparent contrast of this feature, we verified that excluding the polar cells has a negligible effect on the shell-averaged dust-to-gas ratio.
}

In the variable-$\beta$ case, $\langle\epsilon\rangle$ drops by more than two orders of magnitude at $r\simeq R_{\rm B}$ (Fig.~\ref{fig:1dslice_d_to_g_b1e+03}\added[id=R3]{b}) because the recycling flow outside $R_{\rm B}$ prevents dust from entering \added[id=R3]{the Bondi radius} (Fig.~\ref{fig:slice_xz_rhod_b1e+03}b).
Inside $R_{\rm B}$, $\langle\epsilon\rangle$ continues to decrease inward with a gradient similar to the constant-$\beta$ run because the inner convective properties are similar in both cases.

The gray curve in Fig.~\ref{fig:1dslice_d_to_g_b1e+03}\added[id=R3]{b} shows the constant low-$\beta$ run ($\beta_0=10^0$), enabling a direct comparison to \citetalias{kuwahara2026multi_iso}.
In \citetalias{kuwahara2026multi_iso}, the envelope was nearly isothermal and convectively stable everywhere, and it was largely isolated from the recycling flow.
Although the low-$\beta$ and variable-$\beta$ cases both feature an isolated inner envelope and thus show comparable depletion near $R_{\rm B}$, the deep interior differs markedly: $\langle\epsilon\rangle$ becomes much smaller in the low-$\beta$ case because the nearly isothermal gas density increases exponentially toward small radii \added[id=R3]{(Fig.~\ref{fig:rho_dlnT_dlnp_cumQ}a)}.

Figure~\ref{fig:1dslice_d_to_g_b1e+03}\added[id=R3]{c} shows the results for a smaller Stokes number, ${\rm St}=10^{-4}$.
In the constant-$\beta$ case, $\langle\epsilon\rangle$ remains comparable to the ${\rm St}=10^{-3}$ result (blue curves) because dust is tightly coupled to the gas in both cases.
In the variable-$\beta$ case, however, depletion is stronger for ${\rm St}=10^{-4}$ ($\langle\epsilon\rangle \lesssim 10^{-5}$): smaller grains are more efficiently carried by the recycling flow and thus avoid entering the isolated envelope (orange curve).

The trend, namely, that $\langle\epsilon\rangle$ is higher in the non-isolated envelope than in the isolated one, also persists for larger Stokes numbers (Fig.~\ref{fig:1dslice_d_to_g_b1e+03}\added[id=R3]{a}; ${\rm St}=10^{-2}$).
At larger St, dust grains penetrate the envelope more efficiently, so that even isolated envelopes attain higher $\langle\epsilon\rangle$ than in cases with smaller St.

%-------------------------------------------------------------------------------------------
\subsection{Dust dynamics in convective envelopes}\label{sec:Dust dynamics in convective envelopes}
\added[id=R1]{In the following, we distinguish between dust supply from the disk, controlled by envelope isolation, and dust retention within the envelope, controlled by the competition between convection and settling.}
In non-isolated envelopes, dust can enter the Bondi region from the disk, whereas in isolated envelopes, the inner convective region is shielded from incoming dust.
Figure~\ref{fig:slice_xy_dtog_b1e+03} shows snapshots of gas and dust streamlines together with the dust-to-gas ratio in the midplane.
In the constant-$\beta$ (non-isolated) case, convective inflows and outflows create an azimuthal asymmetry in dust penetration and hence in the midplane dust-to-gas ratio (Fig.~\ref{fig:slice_xy_dtog_b1e+03}a).
In the variable-$\beta$ (isolated) case, the dust-to-gas ratio is reduced throughout the Bondi region (Fig.~\ref{fig:slice_xy_dtog_b1e+03}b).
Moreover, the midplane outflow associated with the 3D recycling flow removes dust from the planet vicinity, producing strong depletion even in the second and fourth quadrants in Fig.~\ref{fig:slice_xy_dtog_b1e+03}b.

\added[id=R1]{In the subsequent sections, we investigate dust transport within the envelope. 
We identify two regimes: the settling-dominated and the convection-dominated regime.
In the settling-dominated regime (Sect.~\ref{sec:Settling-dominated regime}), dust settling dominates convective transport in the mean.
In the convection-dominated regime (Sect.~\ref{sec:Convection-dominated regime}), where the terminal-velocity approximation is expected to break down, the dust grains are trapped within the convective circulation.
This occurs for sufficiently small grains (e.g., $s\lesssim0.01$ cm).
}

\begin{figure}[tp]
    \centering
    \includegraphics[width=1\linewidth]{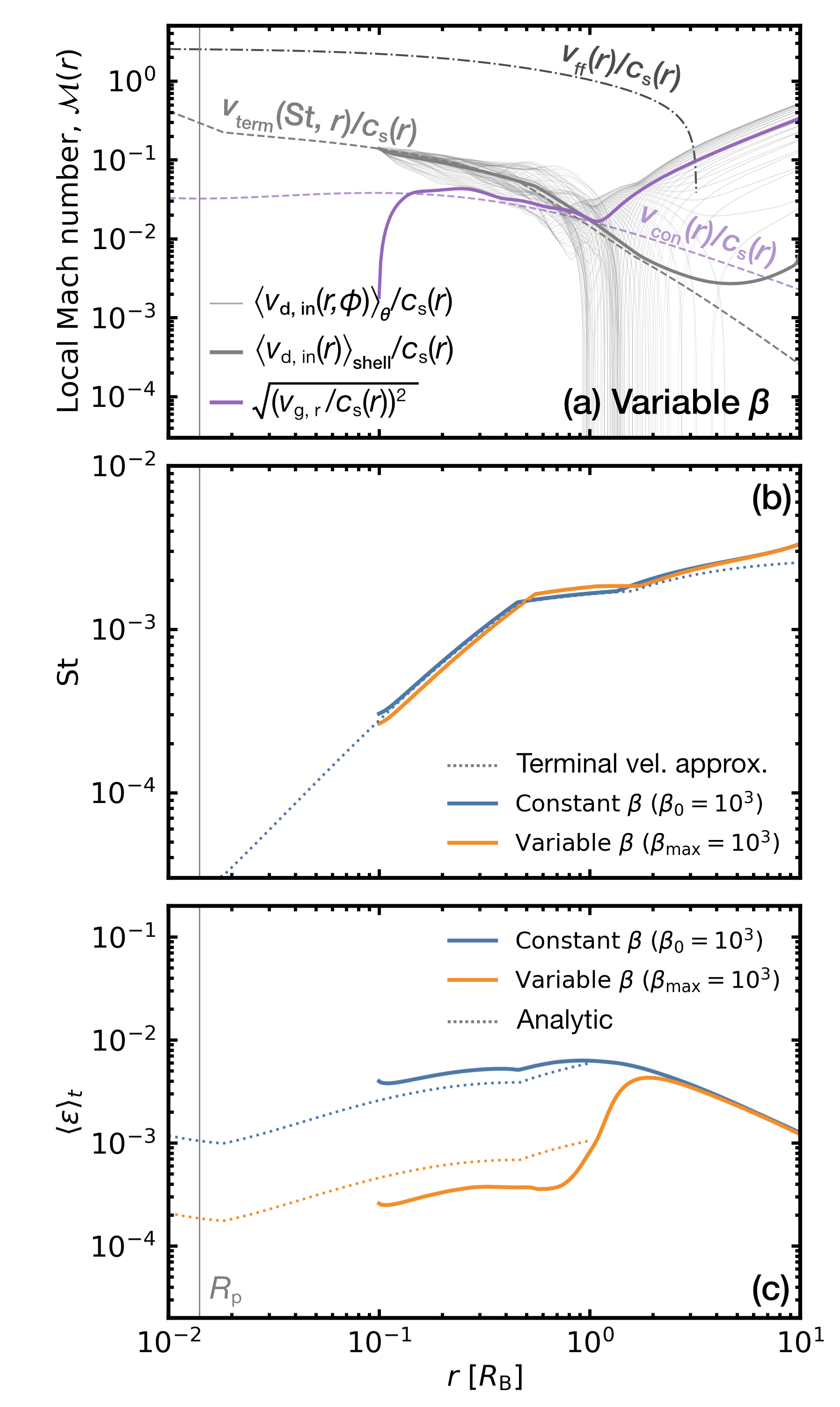}
    \caption{
        Radial profiles of the shell-averaged local Mach number, the Stokes number, and \added[id=R3]{the steady-state} dust-to-gas ratio \added[id=R4]{time-averaged over the last ten orbits} (\textit{top to bottom}) for fixed-size \added[id=R3]{particles with $s=1$ cm (settling-dominated regime).
        In panel~(a), the dot-dashed black and dashed purple curves correspond to Eqs.~\ref{eq:v ff} and \ref{eq:vcon}, respectively.
        The dashed gray curve in panel~(a) and the dotted curve in panel~(b) are obtained by iteratively solving Eqs.~\ref{eq:stopping time} and \ref{eq:vterm fixed size}, combined with a passively irradiated disk model (Eq.~\ref{eq:oka disk model}) and the adiabatic gas density profile (Eq.~\ref{eq:rhog ad}).}
        In the bottom panel, the dotted curve is given by Eq.~\ref{eq:epsilon model} (Sect.~\ref{sec:1D analytic model}).
        \added[id=R3]{The vertical line marks the core radius (Eq.~\ref{eq:core radius}).}
        }
    \label{fig:1dslice_fixed_size_b1e+03_s1e+00cm}
\end{figure}

\begin{figure}[tp]
    \centering
    \includegraphics[width=1\linewidth]{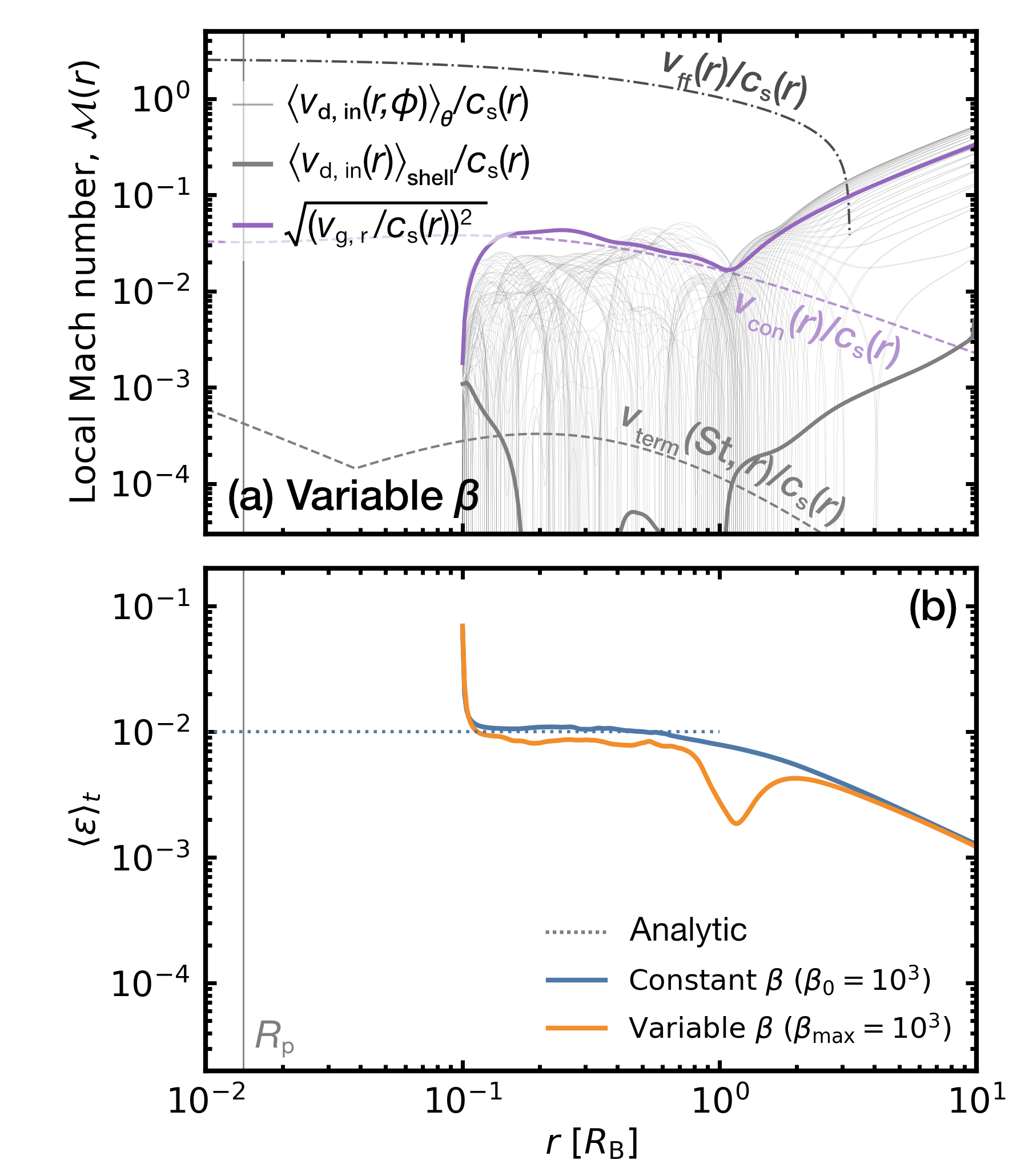}
    \caption{Same as \added[id=R4]{Figs.\,\ref{fig:1dslice_fixed_size_b1e+03_s1e+00cm}a and c}, but for $s=0.01$\,cm \added[id=R3]{(convection-dominated regime)}. The characteristic convective velocity exceeds the terminal velocity everywhere, such that the dust particles are fully entangled with the convective flow. A sharp increase appears at the inner boundary, which shifts inward for a smaller  $r_{\rm in}$ (Appendix~\ref{sec:Convergence tests}).}
    \label{fig:1dslice_fixed_size_b1e+03_s1e-02cm}
\end{figure}

\subsubsection{Settling-dominated regime}\label{sec:Settling-dominated regime}
\added[id=R1]{Dust continues to settle and eventually accretes onto the planet when the \added[id=R3]{shell-averaged} radial motion is settling-like, even though individual grain motions can be strongly perturbed by convection.
All results in \citetalias{kuwahara2026multi_iso} fall in this regime because convection is absent.}
Figure~\ref{fig:1dslice_local_mach_number_vertical} compares (i) \added[id=R4]{polar-angle-averaged} the dust infall velocity at different azimuths, (ii) the shell-averaged dust infall velocity, and (iii) the root-mean-squared radial gas velocity, all normalized by the local sound speed.
For reference, we also plot the terminal, characteristic convective, and free-fall velocities in Fig.~\ref{fig:1dslice_local_mach_number_vertical}.
The terminal and free-fall velocities in code units are given by \citepalias[Eqs.~23 and 24 of][]{kuwahara2026multi_iso}
\begin{align}
    &v_{\rm term}({\rm St},r)=\frac{m\,{\rm St}(r)}{(r/H_{\rm g,0})^2}\,c_{\rm s,0},\label{eq:vterm}\\
    &v_{\rm ff}(r)=\sqrt{\frac{2m}{r/H_{\rm g,0}}\Bigg(1-\frac{r}{R_{\rm H}}\Bigg)}\,c_{\rm s,0}\label{eq:v ff}.
\end{align} 
\added[id=R3]{We note that the expression for $v_{\rm term}$ assumes linear drag, for which the stopping time is independent of the relative velocity. 
In the nonlinear regime, this assumption breaks down. 
We revisit this point in this section (Eq.~\ref{eq:vterm fixed size}) and in Sect.~\ref{sec:1D analytic model}.}

The characteristic convective speed in code units is estimated from mixing-length theory as \added[id=R3]{given in \cite{KL26},} 
\begin{align}
    v_{\rm con}(r,l_{\rm m},\tilde{L}_{\rm acc})\equiv\Bigg(\frac{(\gamma-1)/\gamma}{8\pi (r/H_{\rm g,0})^3}\frac{(l_{\rm m}/R_{\rm B})\tilde{L}_{\rm acc}}{\rho_{\rm g, ad}(r)/\rho_{\rm g,0}}\Bigg)^{1/3}c_{\rm s,0},\label{eq:vcon}
\end{align}
where $l_{\rm m}$ is the mixing length.
For a fully convective envelope ($l_{\rm m}=R_{\rm B}$), Eq.~\ref{eq:vcon} yields $v_{\rm con}\simeq 0.01$--$0.1\,c_{\rm s,0}$, consistent with the numerical results (\added[id=R4]{solid~purple curves} in Fig.~\ref{fig:1dslice_local_mach_number_vertical}).

\added[id=R1]{We confirm that the dust continues to fall when $v_{\rm term}$ exceeds $v_{\rm con}$ \added[id=R4]{everywhere} within the envelope, and we find that the dust infall velocity is well described by $v_{\rm term}$ for the Stokes numbers we considered.}
The thin \added[id=R4]{gray} curves in Fig.~\ref{fig:1dslice_local_mach_number_vertical} show that the dust infall velocities fluctuate strongly and can deviate substantially from $v_{\rm term}$.
Nevertheless, even for ${\rm St}=10^{-4}$, for which $v_{\rm term}>v_{\rm con}$ holds in the innermost region, the shell-averaged infall velocity remains close to $v_{\rm term}$, especially at $r\lesssim R_{\rm B}$.
This indicates that the convective flow reaches a statistically steady state, so that upflows and downflows balance on average and convective fluctuations largely cancel in the shell average.

\paragraph{Simulations with fixed-size particles.}
We have presented results for fixed Stokes numbers so far.
We instead consider fixed particle sizes below.
\added[id=R3]{For fixed-size dust grains, the stopping time and the terminal velocity vary within the envelope. 
As in \citetalias{kuwahara2026multi_iso}, the inward dust velocity was taken to be $v_{\rm d,in}(r)=\min(v_{\rm term},v_{\rm ff})$, with $v_{\rm ff}$ given by Eq.~\ref{eq:v ff}. 
In the nonlinear drag regime, $t_{\rm s}$ and $v_{\rm term}$ were evaluated \added[id=R4]{by root finding} at each radius \citepalias[Eq.~33 of][]{kuwahara2026multi_iso},
\begin{align}
    v_{\rm term}^2\,C_{\rm D}(v_{\rm term})
    =
    \frac{8}{3}\frac{\rho_\bullet s}{\rho_{\rm g}(r)}\,\frac{GM_{\rm p}}{r^2}.\label{eq:vterm fixed size}
\end{align}
}To compute the stopping time consistently with \citetalias{kuwahara2026multi_iso}, we assumed that the planet is embedded in a passively irradiated disk \citep{oka2011evolution},
\begin{align}
    &\Sigma_0=2.7\times10^3\,\text{g/cm}^2\,\bigg(\frac{a_{\rm p}}{1\,\text{au}}\bigg)^{-15/14},\,T_0=150\,\text{K}\,\bigg(\frac{a_{\rm p}}{1\,\text{au}}\bigg)^{-3/7}.\label{eq:oka disk model}
\end{align}
\added[id=R4]{We adopted 1 au as a representative planetary location, since the fully convective envelopes we considered are expected to preferentially occur in the inner disk (Appendices~\ref{sec:1D envelope structure} and \ref{sec:Envelope cooling time model}).}
This gives $c_{\rm s,0}=\sqrt{k_{\rm B}T_0/(\mu m_{\rm H})}\simeq7.3\times10^4\,\mathrm{cm\,s^{-1}}$ and $\rho_{\rm g}\simeq\Sigma_0/(\sqrt{2\pi}H_{\rm g,0})\simeq2.9\times10^{-9}\,\mathrm{g\,cm^{-3}}$, with $k_{\rm B}$ the Boltzmann constant. 
We assumed \added[id=R3]{0.01, 0.1, and 1 cm sized dust with $\rho_\bullet=3\,\text{g/cm}^3$, which leads to ${\rm St}\simeq2.8\times10^{-5}$, $2.8\times10^{-4}$, and $2.8\times10^{-3}$} in the initial state, respectively.

We found no qualitative differences between the fixed-St and fixed-size runs as long as the shell-averaged infall velocity of dust remained close to $v_{\rm term}$.
Figure~\ref{fig:1dslice_fixed_size_b1e+03_s1e+00cm} shows the radial profiles of the characteristic velocities normalized by the local sound speed (panel~a), the Stokes number (panel~b), and the shell-averaged dust-to-gas ratio (panel~c) for \added[id=R3]{$s=1$ cm sized dust.}
Because the envelope becomes denser and hotter toward the planet, St decreases inward (Fig.~\ref{fig:1dslice_fixed_size_b1e+03_s1e+00cm}b).
Despite local perturbations through convection, the shell-averaged dust infall velocity remains close to the terminal velocity, reflecting the statistically steady nature of the convective flow (Fig.~\ref{fig:1dslice_fixed_size_b1e+03_s1e+00cm}a).

The dust-to-gas ratio in the fixed-size runs follows the same qualitative behavior as in the fixed-St runs (Fig.~\ref{fig:1dslice_fixed_size_b1e+03_s1e+00cm}c).
In the constant-$\beta$ case, where the envelope is non-isolated, continuous dust supply from the disk maintains $\langle\epsilon\rangle\sim10^{-2}$ inside $R_{\rm B}$.
By contrast, in the variable-$\beta$ case, the envelope is dynamically isolated and becomes dust-depleted owing to shielding by the outer recycling flow.

\added[id=R1]{The fixed-size results and their close agreement with the fixed-St behavior motivates a 1D description in the regime in which the shell-averaged drift remains dominated by settling (Sect.~\ref{sec:1D analytic model}).
However, as we show in Sect.~\ref{sec:Convection-dominated regime}, the terminal-velocity approximation is expected to break down for sufficiently small grains (e.g., $s\lesssim0.01\,\mathrm{cm}$).
}

%-------------------------------------------------------------------------------------------
\subsubsection{Convection-dominated regime}\label{sec:Convection-dominated regime}
For sufficiently small dust grains ($s\lesssim0.01\,\mathrm{cm}$), \added[id=R3]{we found that} $v_{\rm con}>v_{\rm term}$ holds everywhere within the envelope. 
In this regime, convection can prevent dust from settling.
Figure~\ref{fig:1dslice_fixed_size_b1e+03_s1e-02cm}a shows a snapshot of the shell\added[id=R3]{-}averaged infall velocity, which can take either positive or negative values owing to temporal variability in the convective flow.
At the shown time, the shell-averaged velocity becomes negative at $0.2\,R_{\rm B}<r<0.7\,R_{\rm B}$, indicating that small grains are momentarily transported outward by convective motions rather than settling inward.
This outward transport does not imply a systematic loss of dust from the envelope; instead, the grains are dynamically entangled with the convective circulation.
As a result, the dust-to-gas ratio remains nearly constant throughout the Bondi region (Fig.~\ref{fig:1dslice_fixed_size_b1e+03_s1e-02cm}b).

In the variable-$\beta$ run, the inner convective envelope is dynamically isolated from the outer recycling flow and thus shielded from incoming dust.
Nevertheless, the dust grains \added[id=R3]{that were} initially residing within \added[id=R3]{the Bondi radius} are retained by convective motions, \added[id=R3]{resulting in} a nearly constant $\langle\epsilon\rangle$ although the envelope isolation \added[id=R3]{prevents the delivery of fresh grains to the envelope interior} (orange curve in Fig.~\ref{fig:1dslice_fixed_size_b1e+03_s1e-02cm}b).

%------------------------------------------------------------------------------------------

\section{Extended 1D analytic model}\label{sec:1D analytic model}
As shown in Sect.~\ref{sec:Dust dynamics in convective envelopes}, dust transport in convective envelopes is well described by a 1D shell-averaged framework, \added[id=R1]{except for sufficiently small dust grains.}
We extended the analytic model developed in \citetalias{kuwahara2026multi_iso} to the convective envelopes we explored here.
\added[id=R3]{We derived an analytic expression for the dust density, and, together with the gas density profile, obtained the dust-to-gas ratio.}
All expressions are written in the dimensionless units introduced in Sect.~\ref{sec:Code units and initial conditions}, where $H_{\rm g,0}=c_{\rm s,0}=\Omega_0=\rho_{\rm g,0}=1$, and $\rho_{\rm d,0}=0.01\,\rho_{\rm g,0}$.

\begin{figure}[tp]
    \centering
    \includegraphics[width=1\linewidth]{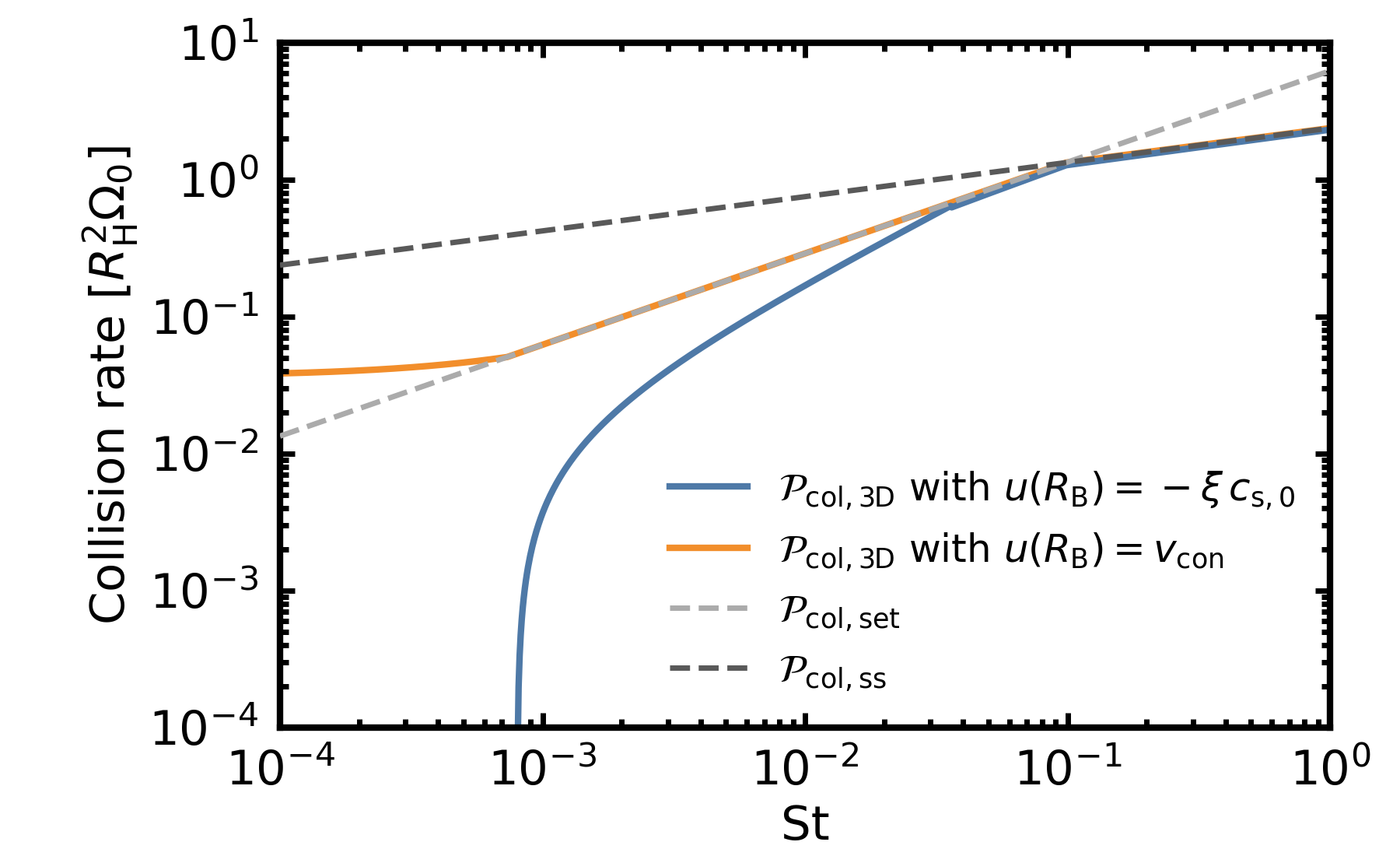}
    \caption{\added[id=R3]{Specific collision rate of pebbles computed from Eq.~\ref{eq:pcol appendix}. The recycling flow associated with the midplane outflow suppresses dust accretion onto the core (blue), whereas the inward component of the convective flow enhances dust accretion (orange). 
    The dashed lines mark the collision rates in the settling regime, depending on the approach velocity of dust (Appendix~\ref{sec:Collision rate of pebbles}).}}
    \label{fig:pcol_p2}
\end{figure}

\added[id=R3]{
When the inward dust mass flux is radially constant, the dust density within the envelope can be expressed as
\begin{align}
    &\rho_{\rm d}(r)=\frac{\dot{M}_{\rm d,acc}}{4\pi r^2v_{\rm d, in}(r)},\label{eq:rhod analytic model}
\end{align}
following \citetalias{kuwahara2026multi_iso}.
Here,
\begin{align}
    \dot{M}_{\rm d,acc}=\sqrt{2\pi}H_{\rm d,0}\rho_{\rm d,0}\,\mathcal{P}_{\rm col,3D}\left({\rm St}|_{r=R_{\rm B}},H_{\rm d,0}\right)\label{eq:Mdot dust}
\end{align}
is the dust accretion rate onto the core, with $\mathcal{P}_{\rm col,3D}$ the specific collision rate of dust grains accounting for the planet-induced non-Keplerian gas flow \citep{okamura2021growth}.
We adopted the collision rate of \citet{okamura2021growth}, but extended it to incorporate
(i) recycling flows associated with midplane outflow (as in \citealt{okamura2021growth} and \citetalias{kuwahara2026multi_iso}) and
(ii) convective motions near $r\simeq R_{\rm B}$ in non-isolated envelopes (Fig.~\ref{fig:pcol_p2}).
Appendix \ref{sec:Collision rate of pebbles} provides the full expressions for the collision rates.}

\added[id=R3]{
The combination of Eq.~\ref{eq:rhod analytic model} with $\rho_{\rm g}(r)$ (Eq.~\ref{eq:rhog ad} or \ref{eq:rho g iso}) provides a closed 1D model for the dust-to-gas ratio.
To account for dust retention by convection, we modeled the dust-to-gas ratio as \added[id=R4]{
\begin{empheq}[left = {\epsilon(r)=\empheqlbrace \,}]{equation}
\begin{aligned}
    &\epsilon_0,
    && \text{if convective and}\\
    &&&  \ 
    v_{\rm con}(r)>v_{\rm term}(r)
    \ \forall\, r\leq R_{\rm B},\\
    &\frac{\rho_{\rm d}(r)}{\rho_{\rm g}(r)},
    && \text{otherwise}.
\end{aligned}
\label{eq:epsilon model}
\end{empheq}
T}he former condition corresponds to a convection-dominated regime in which dust is trapped in convective circulation.
The latter describes a settling-dominated regime, where dust retention by material exchange and filtering by recycling flows are taken into account.
The dotted curves in Figs.~\ref{fig:1dslice_d_to_g_b1e+03}, \ref{fig:1dslice_fixed_size_b1e+03_s1e+00cm}c, and \ref{fig:1dslice_fixed_size_b1e+03_s1e-02cm}b show the resulting predictions, which reproduce the numerical results in the fixed-Stokes and fixed-size runs.
}

\begin{figure}[tp]
    \centering
    \includegraphics[width=1\linewidth]{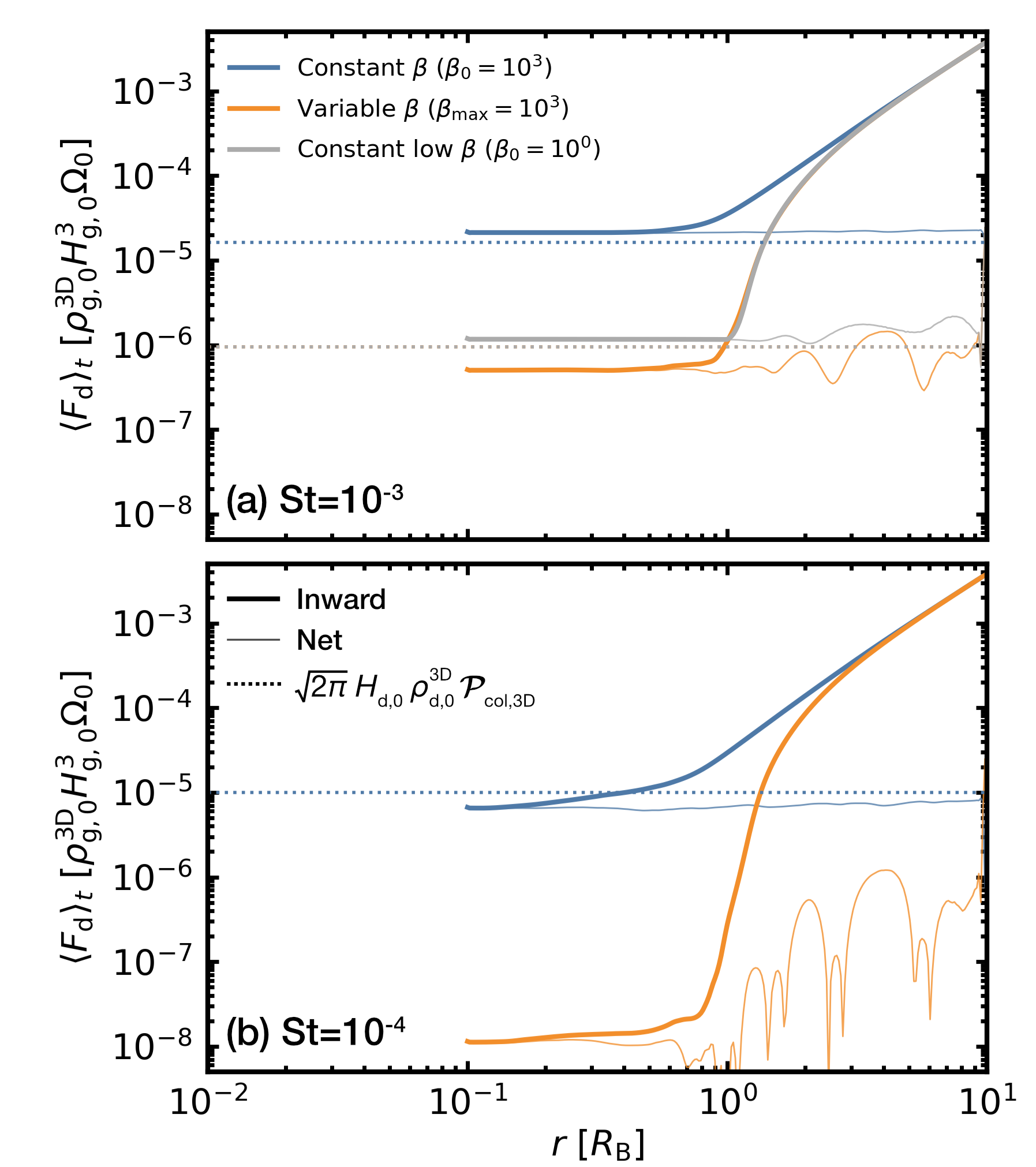}
    \caption{Inward dust mass flux for different $\beta$ prescriptions, \added[id=R2]{which is time-averaged over the last ten orbits.} \added[id=R4]{The thick and thin curves show the inward and net contribution.} The dotted line is given by Eq.~\ref{eq:Mdot dust}, showing \added[id=R2]{the accretion rate of incoming dust}.
    \added[id=R3]{The dust accretion onto the core is enhanced by an order of magnitude in a non-isolated convective envelope (blue) compared to an isolated envelope with a recycling layer (orange and gray).}}
    \label{fig:1dslice_dust_flux_b1e+03}
\end{figure}

\added[id=R4]{The 1D model becomes less accurate for small grains (${\rm St}<10^{-3}$).
For the non-isolated-envelope configuration (constant $\beta$), it overestimates the dust-to-gas ratio (dotted blue curve in Fig.~\ref{fig:1dslice_d_to_g_b1e+03}c) because the model only accounts for inward dust transport, whereas convection also drives outward transport and reduces the net inward flux (Fig.~\ref{fig:1dslice_dust_flux_b1e+03}b).
For the isolated-envelope configuration (variable $\beta$),} the collision-rate model formally yields $\mathcal{P}_{\rm col,3D}=0$, implying perfect dust filtering.
For this reason, the corresponding dotted orange curve is not shown in Fig.~\ref{fig:1dslice_d_to_g_b1e+03}c.
\added[id=R4]{The threshold between dust settling and dust trapping is expected to vary with the accretion luminosity and the extent of the convective region because the characteristic convective velocity scales as $v_{\rm con}\propto (l_mL_{\rm acc})^{1/3}$ (Eq.~\ref{eq:vcon}).
Although we fixed these parameters here, the analytic model can readily be extended to more general envelope structures.
}

%-------------------------------------------------------------------------------------------
%-------------------------------------------------------------------------------------------
\begin{figure*}
    \centering
    \includegraphics[width=1\linewidth]{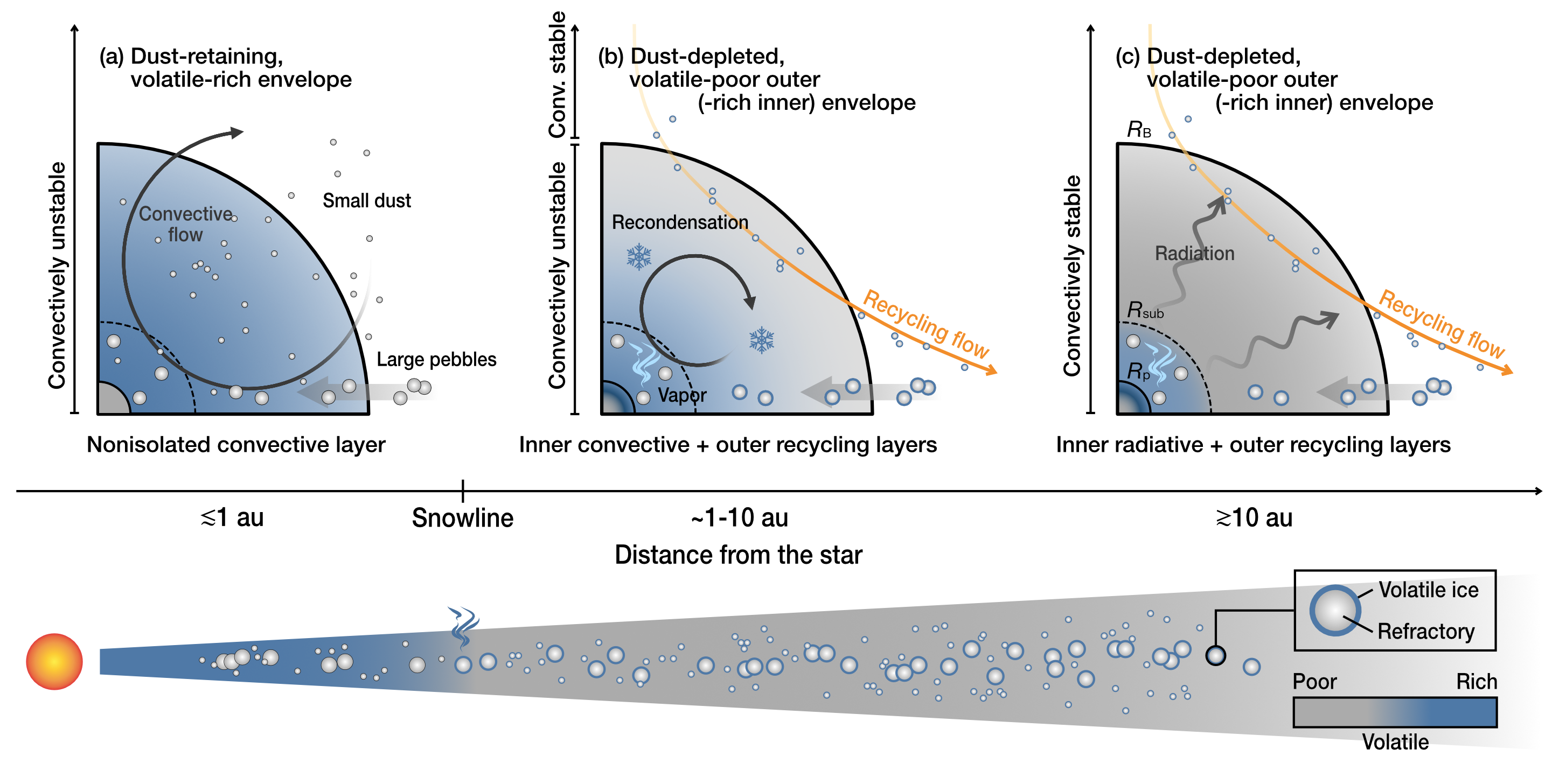}
    \caption{Schematic illustration of (a) a planet with a non-isolated convective envelope, (b) a planet with a convective envelope effectively isolated from an outer recycling flow, and (c) a planet with a radiative envelope effectively isolated from an outer recycling flow. The colors indicate volatile-rich (blue) and volatile-poor (gray) regions. Small dust grains (${\rm St}\lesssim10^{-3}$) are largely coupled to the gas flow, whereas larger pebbles continues to accrete onto the planet independently of the gas dynamics. Volatile species delivered by solids sublimate at the \added[id=R4]{planetary} sublimation front $R_{\rm sub}$, producing volatile vapor. \added[id=R4]{In the nearly isothermal envelopes in the outer disk (panel c), $R_{\rm sub}$ is located deep in the envelope ($<0.1\,R_{\rm B}$) and volatiles can be retained close to the core surface. In contrast, in} \added[id=R1]{convective envelopes, the vapor can be transported outward by convection and may recondense at radii $r>R_{\rm sub}$.}  The planetary core, with radius $R_{\rm p}$, predominantly accretes refractory material. \added[id=R1]{If the inner envelope is dynamically isolated, trapped volatile vapor might contribute to the core bulk composition, yielding a mixture of refractory and volatile material.}}
    \label{fig:schematic_with_disk}
\end{figure*}

\section{Discussions}\label{sec:Discussions}
%-------------------------------------------------------------------------------------------
\subsection{Comparison to previous studies}\label{sec:Comparison to previous studies}
A few studies have investigated dust transport in convective envelopes.
\cite{Popovas:2018a,Popovas:2018b} conducted local 3D hydrodynamic simulations of gas and dust using super-particles and investigated gas and dust dynamics in fully radiative and convective envelopes.
In their setup, the convective envelope was dynamically connected to the surrounding flow, which enables a direct comparison to the non-isolated envelope in this study.
Although their primary aim was to measure particle accretion rates and not the spatial dust distribution inside the envelope, they found that the radially inward dust mass flux is higher in the convective case than in the fully radiative case.
Our constant-$\beta$ (non-isolated) results are consistent with their findings (blue and gray curves in Fig.~\ref{fig:1dslice_dust_flux_b1e+03}).
While \cite{Popovas:2018a} reported systematic inward drift for particles in the size range $s=10^{-3}$--$10^{0}$\,cm,  we identified a convection-dominated regime at sufficiently small grain sizes ($s\lesssim0.01$\,cm), in which grains can become dynamically trapped within the convective circulation (Fig.~\ref{fig:1dslice_fixed_size_b1e+03_s1e-02cm}).

%-------------------------------------------------------------------------------------------
\subsection{Dust processing within envelopes}\label{sec:Dust processing within envelopes}
As in \citetalias{kuwahara2026multi_iso}, we treated solids as either fixed-Stokes or fixed-size particles and neglected dust physics such as growth, fragmentation, ablation, sublimation, and erosion.
Thermal processing is expected to become important only well inside our computational domain, at $r\lesssim0.1\,R_{\rm B}$ (Sect.~6.2 in \citetalias{kuwahara2026multi_iso}).
By contrast, collisional fragmentation can occur within convective envelopes.

We defined the convection-driven mutual collision speed as \citep{ormel2007closed,johansen2020transport}
\begin{align}
    \Delta v_{\rm col}=\sqrt{{\rm St}(r)}\,v_{\rm con}(r).\label{eq:delta v col}
\end{align}
For $s=1$\,cm grains in a passively irradiated disk at 1\,au (Eq.~\ref{eq:oka disk model}), Eq.~\ref{eq:delta v col} gives $\Delta v_{\rm col}\sim1\text{--}10$ m/s, which is comparable to the typical fragmentation thresholds \citep[e.g.,][]{blum2000experiments,wada2013growth}.
A necessary condition for collisions to occur within \added[id=R3]{an envelope} radius $r$ is that the mean collision distance\added[id=R3]{, $l_{\rm col}$,} is shorter than $r$,
\begin{align}
    l_{\rm col}=\frac{(4/3)s\rho_\bullet}{\rho_{\rm d}(r)}\leq r.\label{eq:l col}
\end{align}
The combination of Eq.~\ref{eq:l col} with our analytic estimate of $\rho_{\rm d}(r)$ (Eq.~\ref{eq:rhod analytic model}) yields $l_{\rm col}/r<1$ for $s\lesssim0.1$\,cm grains in parts of the convective envelope, suggesting that fragmentation by mutual collisions can be plausible.
A fully self-consistent treatment therefore requires coupling envelope gas dynamics, dust transport, and dust processing.

\added[id=R3]{One-dimensional approaches of fully convective envelopes that include dust physics argue for the importance of erosion and fragmentation.
\citet{ali2020limits} found that small grains undergo secular infall, but can otherwise be entrained in convective motions and that the dust-to-gas ratio can rise to $\epsilon\sim0.4$ in steady state from an initial value of $\epsilon=0.01$.
They attributed this enhancement to the production of tiny fragments ($s\gtrsim10^{-4}$\,cm) through erosion that are efficiently trapped by convection.
\citet{johansen2020transport} investigated dust transport including dust growth, erosion, and fragmentation, and reported that convection can inhibit dust settling by reducing grain sizes through fragmentation and by lofting grains via convective motions.
For their $1\,M_\oplus$ planet at 1\,au, they found $\epsilon\sim10^{-3}$, which radially increases by factor of approximately 3 at $0.1\,R_{\rm B}\lesssim r\lesssim R_{\rm B}$, and then radially decreases at $r\lesssim0.1\,R_{\rm B}$.
We did not observe an increase in $\epsilon$ within $R_{\rm B}$ like this in our fixed-size calculations, highlighting the role of dust processing, particularly fragmentation and size evolution, in regulating dust transport within the envelope.
}

%-------------------------------------------------------------------------------------------
\subsection{Implications for planet formation and observations}\label{sec:Implications for planet formation and observations}
We discuss implications for planet formation and exoplanet observations below by combining the results of \citetalias{kuwahara2026multi_iso} with those of this study.
Figure~\ref{fig:schematic_with_disk} schematically illustrates the expected envelope structures as a function of orbital distance.
We expect that non-isolated convective envelopes that remain convectively unstable even beyond $R_{\rm B}$ form in hotter inner-disk regions ($\lesssim1$\,au; panel~a), whereas envelopes with an isolated inner layer and an outer recycling layer form in colder outer-disk regions, $\gtrsim1$--$10$\,au (panels~b and~c; Appendix~\ref{sec:1D envelope structure}).

\subsubsection{\added[id=R3]{Planets forming in inner disks ($\lesssim1$\,au)}}
The non-isolated dust-retaining envelopes are expected to form (Fig.~\ref{fig:schematic_with_disk}a).
Because the envelope continuously exchanges material with the disk, the dust content and the gas composition are expected to remain close to those of the surrounding disk gas.
If planets form inside the snowline, the envelope can become volatile rich, whereas the core preferentially accretes refractory material and is comparatively volatile poor.
High dust-to-gas ratios imply high opacities, which help sustain convection and reduce radiative cooling.
The resulting slow cooling can also delay the onset of runaway gas accretion \citep[e.g.,][]{stevenson1982formation,ikoma2000formation}, potentially favoring the formation of inner super-Earths and mini-Neptunes.

\subsubsection{\added[id=R3]{Planets forming in outer disks ($\gtrsim1$--$10$\,au)}}
In outer disks, envelopes are expected to become depleted in small grains, while their interiors can be enriched in volatiles delivered by larger pebbles (Figs.~\ref{fig:schematic_with_disk}b and c).
Volatile species sublimate from accreting pebbles at a radius $R_{\rm sub}$.
The resulting vapor is mixed within a convective layer if present (Fig.~\ref{fig:schematic_with_disk}b), or is trapped inside the sublimation front if convection is absent (Fig.~\ref{fig:schematic_with_disk}c).
\added[id=R4]{We note that in the nearly isothermal envelopes considered in \citetalias{kuwahara2026multi_iso}, the sublimation front was expected to lie deeper than the simulated domain, where the isothermal approximation is likely to break down (Appendix~\ref{sec:1D envelope structure}).}
The outermost envelope, which is continually replenished by recycling flows, \added[id=R4]{remains} closer in composition to the ambient disk gas.
The reduced dust content and opacity can facilitate the early onset of runaway gas accretion, potentially aiding the formation of outer gas giants.

\subsubsection{\added[id=R3]{JWST observations}}
Recent observations with the \textit{James Webb} Space Telescope (JWST) have revealed super- and substellar atmospheric metallicities \added[id=R3]{for} the abundance of elements heavier than helium \citep{taylor2023awesome,fournier2024near,fournier2025transmission,smith2024combined,meech2025bowie,davenport2025toi,liu2025unveiling}.
A pebble-accretion-based population synthesis model attributes this diversity to the formation location and the efficiency of envelope mixing \citep{ohno2026dichotomy}:
planets forming inside the snowline with efficient mixing yield super-stellar metallicities, whereas planets forming outside the snowline with inefficient mixing yield substellar metallicities.
The predicted composition of the outer observable envelope illustrated in Fig.~\ref{fig:schematic_with_disk} lends additional support to this scenario.

\subsection{Model limitations}\label{sec:Model limitations}
The \added[id=R3]{above discussion} primarily concerns the primordial compositions of planetary cores and envelopes.
A more complete picture requires self-consistent hydrodynamic simulations that couple gas dynamics with the thermodynamic effects of \added[id=R3]{volatile sublimation} \citep{wang2023atmospheric}, as well as models that follow the long-term evolution of planetary interiors and atmospheres \citep[e.g.,][]{vazan2023rocky,vazan2024planets}.
In addition, the sustainability of convection and the efficiency of volatile mixing remain uncertain.
We assumed a fixed cooling time and a prescribed accretion luminosity to maintain convection.
In reality, both quantities depend on the evolving dust distribution and the dust accretion rate \added[id=R3]{\citep{Lee:2014,Lee:2015,brouwers2021planets,krapp2024thermodynamic}}.

\added[id=R4]{Another limitation is that we did not solve radiative transfer explicitly.
Analytic studies have shown that the envelope thermal structure is controlled by the cooling efficiency, which depends on the opacity, accretion luminosity, and disk environment \citep[e.g.,][]{rafikov2006atmospheres,piso2014minimum}.
Radiation-hydrodynamic simulations further demonstrated that planetary envelopes develop multidimensional structures, including convective, radiative, and recycling layers, whose extent varies with these thermodynamic conditions \citep{DAngelo:2013,Cimerman:2017,Lambrechts:2017,moldenhauer2021steady,zhu2021global,bailey2024growing}.
At the same time, dust retention depends on the particle size, while the resulting dust distribution modifies the opacity, and hence, the local cooling time.
Capturing this coupled feedback ultimately requires models that simultaneously evolve gas dynamics, dust populations, opacity, and radiative transfer, rather than prescribing $\beta$ a priori.}

%-------------------------------------------------------------------------------------------
%-------------------------------------------------------------------------------------------
\section{Conclusions}\label{sec:Conclusions}
We performed 3D multifluid (gas and dust) simulations to quantify the dust distribution within the envelope of an Earth-mass planet embedded in a protoplanetary disk.
In contrast to \citetalias{kuwahara2026multi_iso}, who assumed nearly isothermal and convectively stable envelopes, we considered nearly adiabatic convectively unstable envelopes.

We found that convectively unstable envelopes can either retain dust or become depleted in dust, depending on (i) whether the inner convective region remains connected to the disk and is thus resupplied with dust, and (ii) whether convective stirring can offset dust settling.
When the convective layer extends beyond the Bondi radius, the envelope remains non-isolated and continuously exchanges material with the background disk, thereby maintaining a dust-rich envelope. 
Such non-isolated convective envelopes are expected to be favored in hotter inner-disk environments (typically $\lesssim \,1\mathrm{au}$).

In contrast, when the envelope is fully convective only inside the Bondi radius while the exterior is convectively stable, a recycling layer develops near the outer envelope edge and shields the inner envelope from incoming solids.
Isolation therefore enables dust depletion by suppressing dust resupply; the subsequent evolution is set by the competition between convective stirring and gravitational settling within the convective region.
When the convective speed exceeds the effective settling speed of solids, convection traps the preexisting dust population and the inner envelope retains the dust.
For the parameter range we explored, this condition is typically satisfied for small grains with sizes $\lesssim 0.01,\mathrm{cm}$.
Otherwise, larger grains settle toward the core largely independently of the convective motion, producing a dust-depleted inner envelope.
Such isolated envelopes are expected to be more common in colder outer-disk regions \added[id=R3]{\citep[typically $\gtrsim 1$–$10\,\mathrm{au}$;][]{KL26}}.

\added[id=R3]{The dust accretion rate can differ by orders of magnitude depending on the envelope structure: recycling suppresses accretion, whereas inward convection enhances it (Fig.~\ref{fig:pcol_p2}).
We therefore extended the 1D analytic model developed in \citetalias{kuwahara2026multi_iso}, originally only applicable to convectively stable envelopes, to the convectively unstable regime. The simulated dust-to-gas profiles agree well.}

These results have direct implications for the primordial composition of planet–envelope systems \added[id=R3]{(Fig.~\ref{fig:schematic_with_disk})}.
Inside the snow line, non-isolated convective exchange causes the envelope to inherit the background disk composition, while the core preferentially accretes refractory solids, yielding a volatile-poor core and a relatively volatile-rich envelope.
Outside the snow line, sublimation of volatiles delivered by large solids can enrich the deep envelope, whereas whether the observable outer envelope becomes volatile enriched depends on the efficiency of convective mixing in the inner envelope.
Our framework can therefore help to interpret trends in atmospheric metallicity of exoplanets inferred from JWST.

Together with \citetalias{kuwahara2026multi_iso}, this work reinforces that the coupled evolution of gas and dust must be modeled to understand dust transport and compositional outcomes in envelopes of disk-embedded planets. 
\added[id=R4]{The idealized envelope structures we considered in this series provide useful end-member cases for isolating how the envelope dynamics regulates dust transport.}
Future models should incorporate gas thermodynamics and dust physics in a unified manner, including radiative cooling, phase changes, grain growth, and fragmentation.

%\section*{Data Availability}

%-------------------------------------------------------------------------------------------
\begin{acknowledgements}
\added[id=R4]{We thank the anonymous referee for an exceptionally thorough and insightful review, whose detailed comments substantially improved the quality of this manuscript.}
We thank the Athena++ developers. The Tycho supercomputer hosted at the SCIENCE HPC center at the University of Copenhagen was used for supporting this work. M.L. acknowledges the ERC starting grant 101041466-EXODOSS. We thank Ziyan Xu for helpful discussions and insightful comments on the radially varying cooling time, $\beta$.
\end{acknowledgements}
%

%-------------------------------------------------------------------------------------------

%% references
%%\raggedright              %% only for adsaa with dvips, not for pdflatex
%\input{accretion_heating.bbl} 
%\ref{Ref}
{\small
\setlength{\bibsep}{0pt}
\bibliography{Ref_aanda}
}

\clearpage
\begin{appendix}

%------------------------------------------
%------------------------------------------
\def\thesection{A}
\setcounter{equation}{0}
\def\theequation{A.\arabic{equation}}
\setcounter{figure}{0}
\def\thefigure{A.\arabic{figure}}

%------------------------------------------
\section{1D envelope structure}\label{sec:1D envelope structure}

\begin{figure}[tp]
    \centering
    \includegraphics[width=1\linewidth]{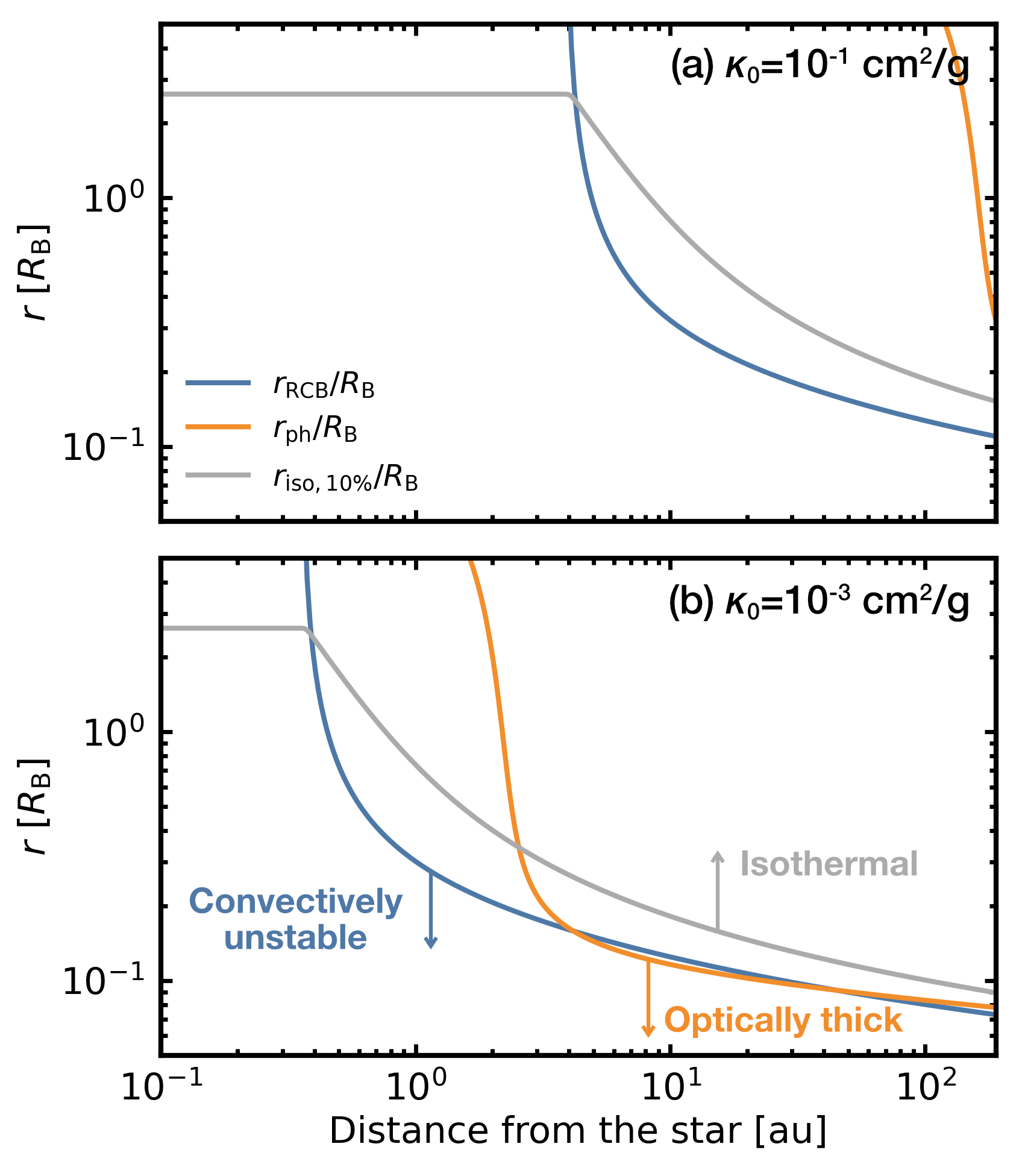}
    \caption{\added[id=R4]{Locations of the RCB, photosphere, and quasi-isothermal radius, normalized by the Bondi radius, as a function of orbital distance for a planet with $m=0.1$. We set $L=10^{27}\,\mathrm{erg/s}$.}}
    \label{fig:characteristic_radii}
\end{figure}

\added[id=R4]{
To relate the adopted $\beta$ prescriptions to physical envelope conditions, we computed 1D envelope structures following \citep[e.g.,][]{rafikov2006atmospheres}
\begin{empheq}[left = {\empheqlbrace \,}]{alignat = 2}
    &\frac{\mathrm{d}p}{\mathrm{d}r}=-\rho_{\rm g}\frac{GM_{\rm p}}{r^2},\label{eq:1D envelope eq1}\\
    &\frac{\mathrm{d}T}{\mathrm{d}r}=\nabla\frac{T}{p}\frac{\mathrm{d}p}{\mathrm{d}r},\label{eq:1D envelope eq2}\\
    &\nabla=\min(\nabla_{\rm rad},\nabla_{\rm ad}),\label{eq:1D envelope eq3}\\
    &\nabla_{\rm rad}=\frac{3\kappa pL}{64\pi\sigma_{\rm SB}GM_{\rm p}T^4},\label{eq:1D envelope eq4}\\
    &\kappa=\kappa_0\,(T/T_0)^2.\label{eq:1D envelope eq5}
\end{empheq}
We assume a dust opacity law of the form $\kappa\propto T^2$ \citep{bell1994using}.
These equations are solved subject to the boundary conditions $T(R_{\rm out})=T_0$, $p(R_{\rm out})=\rho_{\rm g,0}k_{\rm B}T_0/\mu$, and $\rho_{\rm g}(R_{\rm out})=\rho_{\rm g,0}$.
We set $R_{\rm out}=20\,R_{\rm B}$ \citep{rafikov2006atmospheres}.
As in the main text, we adopt a passively irradiated disk model as the background disk (Eq.~\ref{eq:oka disk model}) and $L=10^{27}\,\mathrm{erg/s}$.}

\added[id=R4]{Figure~\ref{fig:characteristic_radii} shows the locations of the RCB, photosphere, and quasi-isothermal radius as a function of orbital distance for the fiducial planet mass, $m=0.1$.
The photosphere is defined by $\tau(r_{\rm ph})=\int_r^{R_{\rm out}}\kappa\rho_{\rm g}\,\mathrm{d}r=1$, while the quasi-isothermal radius is defined by $|(T(r_{\rm iso,10\%})-T_0)/T_0|=0.1$.
For a high-opacity case ($\kappa=10^{-1}\,\mathrm{cm^2\,g^{-1}}$; Fig.~\ref{fig:characteristic_radii}a), the entire envelope, including regions outside the Bondi radius, becomes convectively unstable in the inner disk ($\lesssim5$ au).
In contrast, for a low-opacity case ($\kappa=10^{-3}\,\mathrm{cm^2\,g^{-1}}$; Fig.~\ref{fig:characteristic_radii}b), the RCB, photosphere, and quasi-isothermal radius shift inward and lie at $r\lesssim0.1\,R_{\rm B}$ beyond $\sim10$ au.
Therefore, the nearly isothermal envelopes studied in \citetalias{kuwahara2026multi_iso} are most applicable to low-opacity envelopes in the outer disk, whereas the fully convective envelopes considered here are broadly representative of high-opacity envelopes in the inner disk.
We emphasize that these 1D calculations are intended only to provide physical guidance for interpreting the adopted $\beta$ prescriptions.
In reality, the 3D envelope structure is further shaped by recycling flows, convection, shear, and time-dependent energy transport.}

%-------------------------------------------------------------------------------------------
%-------------------------------------------------------------------------------------------
\def\thesection{B}
\setcounter{equation}{0}
\def\theequation{B.\arabic{equation}}
\setcounter{figure}{0}
\def\thefigure{B.\arabic{figure}}

\section{Envelope cooling time model}\label{sec:Envelope cooling time model}

\begin{figure}[tp]
    \centering
    \includegraphics[width=1\linewidth]{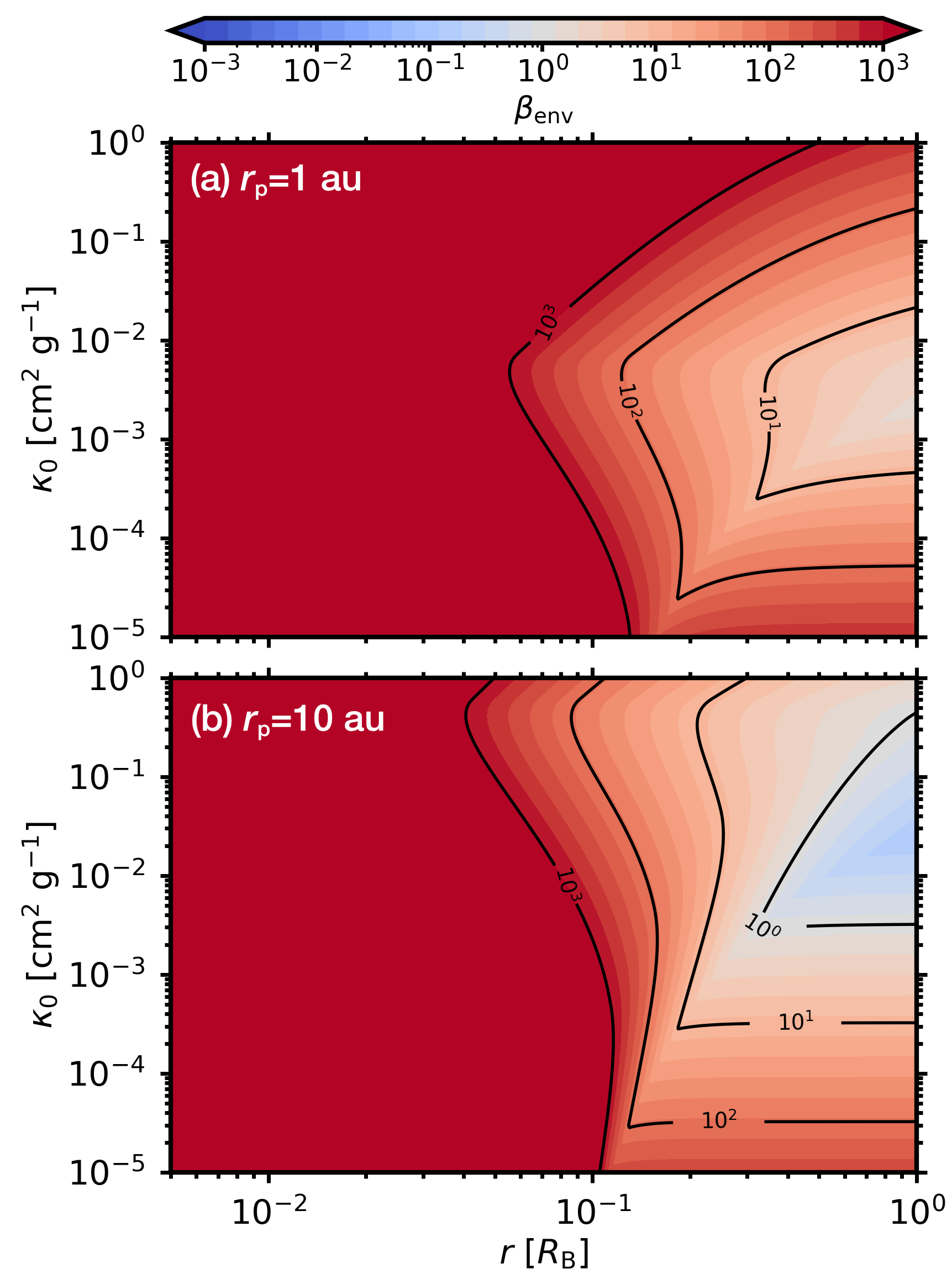}
    \caption{\added[id=R4]{Contours of the estimated envelope cooling parameter, $\beta_{\rm env}$, for a planet with $m=0.1$.}}
    \label{fig:envelope_beta_contour}
\end{figure}

\added[id=R4]{
Using the envelope structures described above, we estimated the local thermal relaxation time by considering optically thin and optically thick limits}\footnote{This estimation is different from the cooling time analysis in \cite{KL26}. \cite{KL26} estimated the cooling time by combining the cooling time of the background disk gas---which constrains the cooling time at the outer edge of the envelope---with the depth-dependent cooling time derived from a two-layer envelope model \citep{piso2014minimum}.} \added[id=R4]{\citep[e.g.,][]{spiegel1957smoothing,Lin:2015,barranco2018zombie},
\begin{align}
    &t_{\rm thin}=\frac{c_{\rm V}}{16\,\kappa(T)\sigma_{\rm SB}T(r)^3},\\
    &t_{\rm thick}=\frac{3\,\kappa(T)\rho_{\rm g}(r)^2c_{\rm V}\ell_{\rm th}^2}{16\sigma_{\rm SB}T(r)^3},
\end{align}
where $c_{\rm V}$ is the heat capacity and $\ell_{\rm th}$ is the characteristic length scale of the thermal perturbation.
We adopt $\ell_{\rm th}=R_{\rm B}$ and define
\begin{align}
    &t_{\rm relax}=\max(t_{\rm thin},t_{\rm thick}),\quad\beta_{\rm env}=\Omega_{\rm K}(r_{\rm p})\,t_{\rm relax},
\end{align}
where $\Omega_{\rm K}(r_{\rm p})$ is the Keplerian frequency at the planet's orbital location.
}

\added[id=R4]{
The cooling-time estimate supports interpreting the constant-$\beta$ model ($\beta_0=10^3$) as a representation of high-opacity envelopes in the inner disk.
For $\kappa_0=10^{0}\ {\rm cm^2\,g^{-1}}$ at 1 au, the local estimate gives $\beta_{\rm env}\gtrsim10^3$ throughout most of the envelope (Fig.~\ref{fig:envelope_beta_contour}a).
Although $\beta_{\rm env}$ can exceed $10^3$, the envelope is already close to the adiabatic limit in this regime.
We therefore do not expect different gas dynamics at larger $\beta$ \citep{KL26}.}

\added[id=R4]{Likewise, the cooling-time estimate supports the applicability of the constant low-$\beta$ model ($\beta_0=1$) to low-opacity envelopes in the outer disk.
For $\kappa_0=10^{-3}\ {\rm cm^2\,g^{-1}}$ at 10 au, the estimate gives $\beta_{\rm env}$ of order unity outside approximately $0.1\,R_{\rm B}$ (Fig.~\ref{fig:envelope_beta_contour}b), corresponding to the region primarily explored in our simulations.}

\added[id=R4]{
The variable-$\beta$ model was designed to approximate envelopes containing a slowly cooling inner region and a more rapidly cooling outer region.
The 1D structure model predicts a convectively unstable inner envelope surrounded by a convectively stable outer region at several au, qualitatively similar to the intended structure of the variable-$\beta$ simulations (Fig.~\ref{fig:characteristic_radii}).
However, the corresponding cooling-time estimate predicts a smooth radial variation of $\beta_{\rm env}$ rather than the step-like transition adopted in our simulations (Fig.~\ref{fig:envelope_beta_contour}).
No single opacity model therefore reproduces the envelope structure and the adopted cooling prescription simultaneously.
Instead, realistic envelopes are expected to contain a finite radiative layer separating the inner convective region from the outer recycling flow.
Such a three-layer structure has been found in radiation-hydrodynamic simulations \citep[e.g.,][]{Lambrechts:2017} and would naturally produce a much broader transition than assumed here.
The variable-$\beta$ prescription should therefore be regarded as an idealized representation of this transition rather than a self-consistent opacity model.
We emphasize that calibrating a prescribed $\beta$-cooling model against radiative transfer is non-trivial and requires future systematic comparisons with radiation-hydrodynamic simulations.
}

%-------------------------------------------------------------------------------------------
\def\thesection{C}
\setcounter{equation}{0}
\def\theequation{C.\arabic{equation}}
\setcounter{figure}{0}
\def\thefigure{C.\arabic{figure}}

\begin{figure}[tp]
    \centering
    \includegraphics[width=1\linewidth]{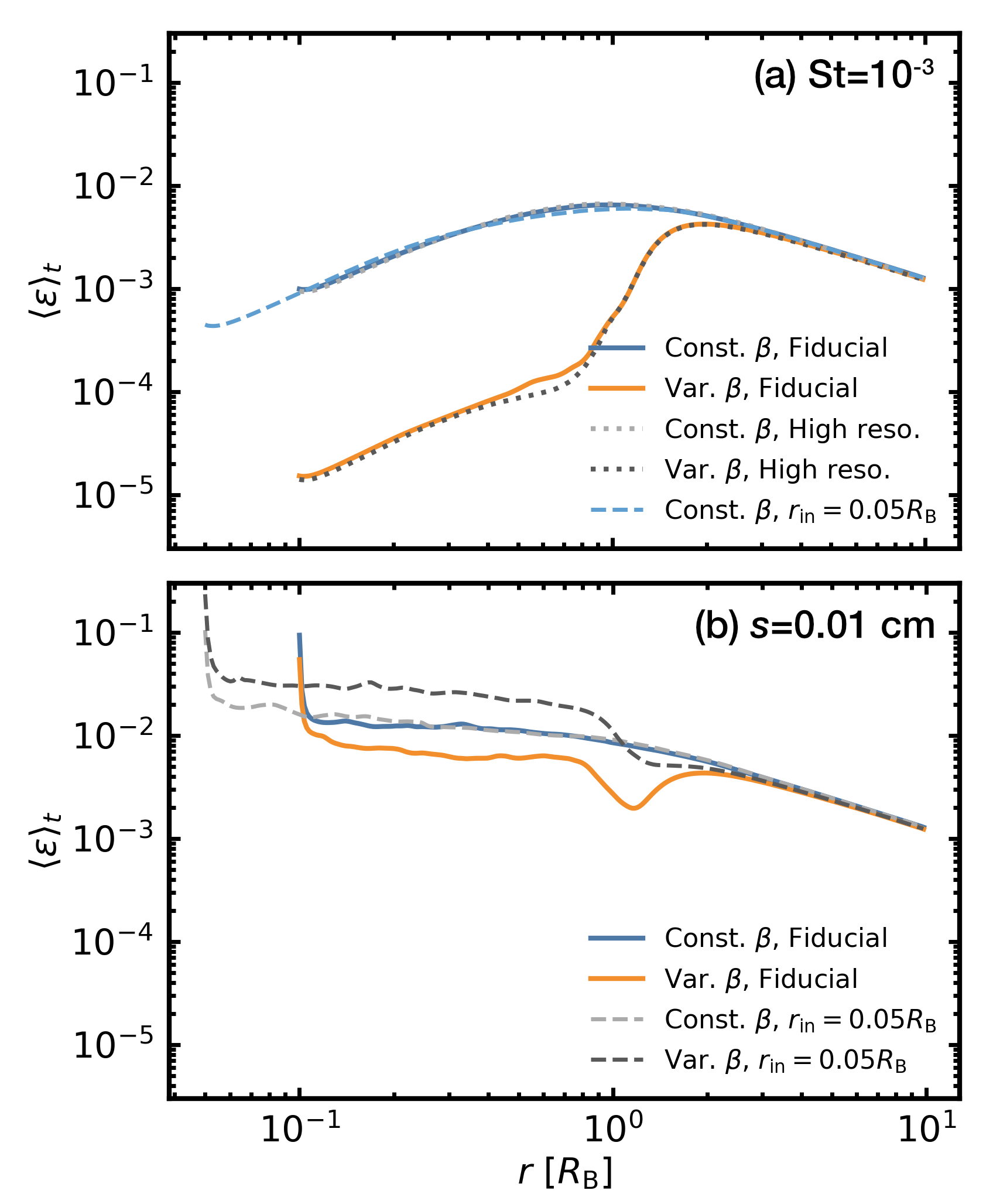}
    \caption{Shell-averaged dust-to-gas ratio for different numerical settings. \added[id=R4]{These curves are also time-averaged over the last 10 orbits}. This figure compares our fiducial setup (solid curves) with the high-resolution runs (dotted) and the run with a smaller $r_{\rm in}$ (dashed).}
    \label{fig:convergence_test_d_to_g_1D_b1e+03}
\end{figure}
%-------------------------------------------------------------------------------------------

\section{Convergence tests}\label{sec:Convergence tests}
We assess the numerical convergence of our fiducial runs in Fig.~\ref{fig:convergence_test_d_to_g_1D_b1e+03}, where we investigate the dependence of the simulation results on the resolution and the
size of the inner boundary.
The parameter choices adopted for the convergence tests are
summarized in Table~\ref{tab:hydro simulations}.

%-------------------------------------------------------------------------------------------
\def\thesection{D}
\setcounter{equation}{0}
\def\theequation{D.\arabic{equation}}
\setcounter{figure}{0}
\def\thefigure{D.\arabic{figure}}

%-------------------------------------------------------------------------------------------

\section{Specific collision rate of pebbles}\label{sec:Collision rate of pebbles}
Following \citet{okamura2021growth}, we model the \added[id=R2]{collision rate} that pebbles drifting with the background shear are accreted onto a planet in the presence of a planet-induced, non-Keplerian gas flow.
The main extension introduced in this work is the inclusion of convective gas motions at the Bondi radius.

The collision rate is written as either a 2D-limit expression or its 3D extension,
\begin{empheq}[left = {\mathcal{P}_{\rm col}=\empheqlbrace \,}]{alignat = 2}
    &\mathcal{P}_{\rm col,2D}({\rm St}),\nonumber\\
    &\mathcal{P}_{\rm col,3D}({\rm St},H_{\rm d,0}),\label{eq:pcol appendix}
\end{empheq}
and
\tiny
\begin{empheq}[left = {\empheqlbrace \,}]{alignat = 2}
    &\mathcal{P}_{\rm col,3D}({\rm St},H_{\rm d,0})=\Bigg[\bigg(\mathcal{P}_{\rm col,2D}({\rm St})\bigg)^{-2}+\Bigg(\mathcal{P}_{\rm col,2D}({\rm St})\frac{x_{\rm ss}}{0.65\,H_{\rm d,0}}\Bigg)^{-2}\Bigg]^{-1/2},\\
    &\mathcal{P}_{\rm col,set}=3\,(2C_1{\rm St})^{2/3}\,R_{\rm H}^2\Omega_0,\\
    &\mathcal{P}_{\rm col,flow}=2\frac{R_{\rm B}}{R_{\rm H}}\Bigg[\frac{3\,{\rm St}}{(R_{\rm B}/R_{\rm H})^2} + \frac{u(R_{\rm B})}{c_{\rm s,0}}\sqrt{\frac{3}{(R_{\rm B}/R_{\rm H})}}\Bigg]\,R_{\rm H}^2\Omega_0,\\
    &\mathcal{P}_{\rm col,ss}=2\sqrt{6}C_1\Bigg(\frac{\rho_{\bullet}}{\rho_{\rm g}}\frac{c_{\rm s}}{R_{\rm H}\Omega_0}\frac{l_{\rm mfp}}{R_{\rm H}}{\rm St}\Bigg)^{1/4}\,R_{\rm H}^2\Omega_0,\\
    &x_{\rm ss}=
    \begin{cases}
        (2C_1{\rm St})^{1/3}\,R_{\rm H}&\text{if } \mathcal{P}_{\rm col,2D}=\mathcal{P}_{\rm col,set},\\
        2\,R_{\rm B}&\text{if } \mathcal{P}_{\rm col,2D}=\mathcal{P}_{\rm col,flow},\\
        \displaystyle
        (8/3)^{1/4}\sqrt{C_1}\Bigg(\frac{\rho_{\bullet}}{\rho_{\rm g}}\frac{c_{\rm s}}{R_{\rm H}\Omega_0}\frac{l_{\rm mfp}}{R_{\rm H}}{\rm St}\Bigg)^{1/8}\,R_{\rm H}&\text{if } \mathcal{P}_{\rm col,2D}=\mathcal{P}_{\rm col,ss},
    \end{cases}\\
    & C_1 = 1.5.
\end{empheq}
\normalsize
The expression for $\mathcal{P}_{\rm col,2D}$ is given below.

To account for perturbed, non-Keplerian gas flows, we defined $\mathcal{P}_{\rm col,flow}$ in the above, where $u(R_{\rm B})$ parameterizes the characteristic radial gas motion
at the Bondi radius (Sect.~\ref{sec:1D analytic model}),
\begin{empheq}[left={u(R_{\rm B})=\empheqlbrace}]{alignat=2}
    &-\xi\,c_{\rm s,0},\\
    &v_{\rm con}(R_{\rm B},l_{\rm m},\tilde{L}_{\rm acc}).
\end{empheq}
In \citet{okamura2021growth} and \citetalias{kuwahara2026multi_iso},
only the first case, $u(R_{\rm B})=-\xi\,c_{\rm s,0}$, was considered.
With this choice, $\mathcal{P}_{\rm col,flow}$ reduces to
$\mathcal{P}_{\rm col,ho}$ introduced by \citet{okamura2021growth}
(their Eq.~43).
The parameter $\xi$ quantifies the effect of the midplane outflow associated with the recycling flow, on suppressing pebble accretion \citep{Kuwahara:2019}.
While \citet{okamura2021growth} adopted $\xi=0.1\,m$, we use \added[id=R2]{$\xi=0.08\,m$} to better match our numerical results same as in \citetalias{kuwahara2026multi_iso}.

In this work, we additionally consider $u(R_{\rm B})=v_{\rm con}$ to describe inward convective motions reaching $r\simeq R_{\rm B}$, which can enhance the effective \added[id=R2]{collision rate}.
With these definitions, we compute $\mathcal{P}_{\rm col,2D}$ as
\begin{empheq}[left = {\mathcal{P}_{\rm col,2D}=\empheqlbrace \,}]{equation}
\tiny
\begin{aligned}
    &\min\big(\mathcal{P}_{\rm col,flow},\,\mathcal{P}_{\rm col,set},\,\mathcal{P}_{\rm col,ss}\big)
    && \text{if } u(R_{\rm B})=-\xi\,c_{\rm s,0},\\
    &\min\bigg(\max\big(\mathcal{P}_{\rm col,flow},\,\mathcal{P}_{\rm col,set}\big),\,\mathcal{P}_{\rm col, ss}\bigg)
    && \text{if } u(R_{\rm B})=v_{\rm con}\geq v_{\rm d,in}(R_{\rm B}),\\
    &\mathcal{P}_{\rm col,set}
    && \text{if } u(R_{\rm B})=v_{\rm con}< v_{\rm d,in}(R_{\rm B}).
\end{aligned}
\label{eq:Pcol_2D}
\end{empheq}
\normalsize
The first and second cases correspond to (i) an isolated envelope with a recycling flow (convectively stable at $r\geq R_{\rm B}$) and (ii) a non-isolated envelope in which the gas is convectively unstable even outside the Bondi radius, respectively.
When the Stokes number varies with radius, we evaluate
$\mathcal{P}_{\rm col}$ using its value at $r=R_{\rm B}$.
As in \citetalias{kuwahara2026multi_iso}, we (i) ignore the collision rate in the gas-free regime, relevant for ${\rm St}>10^{0}$ (defined as $\mathcal{P}_{\rm atm}$ in \citealt{okamura2021growth}), and (ii) evaluate the collision rate in the supersonic regime, $\mathcal{P}_{\rm col,ss}$, using $\rho_{\rm g}$, $c_{\rm s}$, $\Omega_0$, and $l_{\rm mfp}$ at 1~au in the passively irradiated disk model adopted in Sect.~\ref{sec:Settling-dominated regime}.

Figure~\ref{fig:pcol_p2} shows the resulting \added[id=R2]{collision rate}.
For $u(R_{\rm B})=-\xi\,c_{\rm s,0}$, the midplane outflow associated with the recycling flow reduces $\mathcal{P}_{\rm col}$.
In contrast, for $u(R_{\rm B})=v_{\rm con}$, inward convective motions enhance the \added[id=R2]{collision rate}, yielding
$\mathcal{P}_{\rm col}>\mathcal{P}_{\rm col,set}$.

\end{appendix}
%------------------------------------------
\end{document}